\documentclass[preprint2]{aastex}

\newcommand{\kms}{\mbox{km\,s$^{-1}$}}

\slugcomment{To be submitted to the Astronomical Journal}

\shorttitle{Draco}
\shortauthors{Odenkirchen et al.}

\begin{document}

\title{New insights on the Draco dwarf spheroidal galaxy from SDSS: 
a larger radius and no tidal tails}

\author{
Michael Odenkirchen\altaffilmark{1}, 
Eva K. Grebel\altaffilmark{1}, 
Daniel Harbeck\altaffilmark{1}, 
Walter Dehnen\altaffilmark{1}, 
Hans-Walter Rix\altaffilmark{1}, 
Heidi Jo Newberg\altaffilmark{2},
Brian Yanny\altaffilmark{3},
Jon Holtzman\altaffilmark{4}
Jon Brinkmann\altaffilmark{5},
Bing Chen\altaffilmark{5},
Istvan Csabai\altaffilmark{6},
Jeffrey J.E. Hayes\altaffilmark{5},
Greg Hennessy\altaffilmark{7},
Robert B. Hindsley\altaffilmark{8},
\v{Z}eljko Ivezi\'{c}\altaffilmark{9},
Ellyne K. Kinney\altaffilmark{5},
S.J. Kleinman\altaffilmark{5},
Dan Long\altaffilmark{5},
Robert H. Lupton\altaffilmark{9},
Eric H. Neilsen\altaffilmark{5},
Atsuko Nitta\altaffilmark{5},
Stefanie A. Snedden\altaffilmark{5},
Donald G. York \altaffilmark{10}
}

\altaffiltext{1}{Max-Planck-Institut f\"ur Astronomie, K\"onigstuhl 17, 
D-69117 Heidelberg, Germany; odenkirchen@mpia-hd.mpg.de, 
grebel@mpia-hd.mpg.de}
\altaffiltext{2}{Dept. of Physics, Applied Physics and Astronomy, 
Rensselaer Polytechnic Institute, Troy, NY 12180}
\altaffiltext{3}{Fermi Accelerator National Laboratory, P.O. Box 500, 
Batavia, IL 60510}
\altaffiltext{4}{Department of Astronomy, New Mexico State University, 
Box 30001/Department 4500, Las Cruces, NM 88003-8001}
\altaffiltext{5}{Apache Point Observatory, P.O.Box 59, 2001 Apache 
Point Road, Sunspot, NM 88349-0059}
\altaffiltext{6}{Department of Physics and Astronomy, Johns Hopkins 
University, Charles and 34th Street, Bloomberg Center, Baltimore, MD 
21218-2686}
\altaffiltext{7}{US Naval Observatory, 3450 Massachusetts Avenue, NW, 
Washington, DC 20392-5420}
\altaffiltext{8}{US Naval Research Laboratory, 4555 Overlook Avenue, SW,
Washington, DC 20375-5352}
\altaffiltext{9}{Princeton University Observatory, Peyton Hall, Ivy Lane,
Princeton, NJ 08544-1101}
\altaffiltext{10}{Department of Astronomy and Astrophysics, University 
of Chicago, 5640 South Ellis Avenue, Chicago, IL 60637}

\begin{abstract}
We have investigated the spatial extent and structure of the Draco
dwarf spheroidal galaxy using deep wide-field multicolor CCD photometry 
from the Sloan Digital Sky Survey (SDSS). Our study covers an 
area of 27 square degrees around the center of the Draco dwarf and 
reaches two magnitudes below the level of Draco's horizontal branch.  
The SDSS photometry allows very effective filtering in color-magnitude 
space. With such filtering the density of the foreground of Galactic field 
stars is decreased by more than an order of magnitude and the stellar 
population 
of the Draco dwarf galaxy thus stands out with much higher contrast than 
in former investigations. 
We show that the spatial distribution of Draco's red giants, red horizontal 
branch stars and subgiants down to $i^*$=21.7~mag does not provide evidence 
for the existence of tidally induced tails or a halo of unbound stars. 
The projected surface density of the dwarf galaxy is flattened with a
nearly constant ellipticity of $0.29\pm 0.02$ at position angle 
$88\pm 3^\circ$.  The radial profile can be fit by King models as well as
by a generalized exponential.
Using the empirical King (1962) profile the core radius and the limiting 
(or tidal) radius along the major axis are $r_c = 7\farcm7$ and 
$r_t = 40\farcm1$, respectively; the latter means that the size of the 
Draco dwarf galaxy is 40\% larger than previously measured. 
Fitting the profile of King's (1966) theoretical models yields a still 
larger limiting radius of $r_t = 49\farcm5$. 
There is no clear indication of a tail-like extension of the Draco population 
beyond this radius.
A break in the radial surface density profile which might indicate a
halo of extratidal stars is also not found in our Draco data. 
We conclude that down to the above magnitude limit tidal effects can 
only exist at a level of $10^{-3}$ of the central surface density of Draco
or below. The regular structure of Draco found from the new data argues
against it being a portion of an unbound tidal stream and lends support 
to the assumption of dynamical equilibrium which is the basis for mass 
estimation. 
The changes in the values for the core radius and limiting radius imply 
that the total mass of Draco is higher by more than a factor of two.
Using a King (1966) spherical model of equivalent size as a reference and 
adopting a line-of-sight velocity dispersion of either 10.7~\kms or 8.5~\kms  
(Armandroff et al.\ 1995), we derive estimates of the total mass within 
radius $r_t$ of $3.5\pm 0.7\times 10^7~M_\odot$ and 
$2.2\pm 0.5\times 10^7~M_\odot$, respectively. 
From the combined $i$ band flux of all possible Draco members that lie 
within major axis radius $r_t$ we determine the total luminosity 
of the Draco dwarf galaxy as $(L/L_\odot)_{i} = 2.4\pm 0.5 \times 10^5$. 
This includes corrections for the flux of the foreground stars and the 
unseen fainter part of the Draco population. 
We thus obtain overall mass-to-light ratios $M/L_{i}$ of $146\pm 42$ or 
$92\pm 28$ in solar units. 
In summary, our results strengthen the case for a strongly dark matter 
dominated, bound stellar system.

\end{abstract}


\section{Introduction}

The Milky Way is surrounded by a number of small faint companion galaxies 
that are gravi\-tationally bound to it. 
Nine such Milky Way dwarf satellites at galactocentric distances 
between 16~kpc and 250~kpc are presently known.   
The dwarf spheroidal galaxy in Draco (hereafter called Draco dSph or 
simply Draco), located at a distance of about 71~kpc (see \S~5.1), is one 
of the two least luminous of these companions. 
Estimates of the dynamical masses of the Milky Way dwarf satellites 
based on their spatial extent and internal velocity dispersion lead 
to the conclusion that most of them have high mass-to-light ratios 
(30 - 100) and hence must contain large amounts of dark matter (see 
Mateo 1998 and references therein). 
The parameters of the Draco dSph imply a particularly high mass-to-light 
ratio of $M/L \simeq 60 - 90$ (Armandroff et al. 1995). 
Thus Draco must be completely dominated by dark matter if the 
underlying assumption of dynamical equilibrium holds. \\

However, even with a large fraction of dark matter the total masses 
of dwarf spheroidal galaxies like Draco are at most a few
$10^7$ $M_\odot$ while their spatial extent is several hundred parsecs. 
Therefore these systems are susceptible to perturbations by the tidal 
forces of the Galactic potential if their orbits are sufficiently 
eccentric to bring them to galactocentric distances of about 20 kpc 
or closer. As the result of such perturbations, a dwarf galaxy may lose 
significant amounts of mass and eventually be disrupted. 
This has a number of important implications: \\
1. Tidal mass loss and disruption of dwarf galaxies can shed light on 
the formation of the stellar Galactic halo and its spatial and kinematic 
structure (Johnston et al.\ 1996, Johnston 1998, Helmi \& White 1999). \\
2. The spatial distribution of tidal debris can provide constraints on the 
space motions and orbits of the companions (e.g., Odenkirchen et al.\ 2001). 
These would, e.g., be useful for testing the hypothesis that the Milky Way 
satellites are grouped in certain streams (e.g., Lynden-Bell \& Lynden-Bell 
1995). Traces of the orbits of the dwarf galaxies would also provide 
constraints on the gravitational potential of the Galactic halo
(e.g., Kuhn 1993, Johnston et al.\ 1999a).\\
3. If a dwarf galaxy is tidally perturbed it may be incorrect to estimate 
its mass via the assumption of dynamical equilibrium. Kuhn \& Miller (1989) 
have argued that tidal heating can inflate the velocity dispersion of a dwarf
galaxy and hence lead to overestimated masses and mass-to-light ratios. 
Numerical simulations by Klessen \& Kroupa (1998) have shown that in the 
extreme case of complete disruption the observed properties of local dwarf 
galaxies may under certain conditions be explicable without any dark 
matter. This point however is controversial (see Olszewski 1998 and 
references therein). Piatek \& Pryor (1995) found that even when tides 
affect the structure and the kinematics of a dwarf spheroidal, they fail 
to inflate the central mass-to-light ratio.\\

An impressive example of tidal interaction is given by the Sagittarius dwarf 
spheroidal, the closest companion of the Milky Way which is observed while 
being disrupted in a close encounter with the Galaxy (Ibata et al. 1997). 
Sagittarius has a very elongated and distorted shape and part of its stellar 
content is stretched out in a tidal stream along its orbit (Mateo et al.\ 
1998, Majewski et al.\ 1999, Johnston et al.\ 1999b, Ibata et al.\ 2000). 
This shows that tidal effects can indeed play an important role.
It is therefore desirable to find out if the other (more distant) dwarf 
satellites also reveal signs of tidal mass loss and disruption. 
Majewski et al.\ (2000) showed that the giants of the Carina dwarf 
spheroidal have an extended distribution that reaches beyond the estimated 
tidal limit and produces a break in the projected radial density profile. 
Martinez-Delgado et al.\ (2001) recently reported the detection of a
tidal extension for the dwarf spheroidal in Ursa Minor.   

From recent studies of Draco there seemed to be evidence that this dwarf 
also has an extratidal stellar component.  Irwin \& Hatzidimitriou (1995; 
hereafter IH95) investigated the spatial structure of all known dwarfs 
except Sagittarius by means of star counts from digitized photographic 
plates. 
They found for Draco, as well as for most of the other dwarfs, significant 
overdensities of stars outside the limiting radius of their best-fit King 
profile. 
Smith, Kuhn \& Hawley (1997, hereafter SKH97) reported preliminary results 
of CCD photometry in a number of fields along the major and minor axis of 
the Draco dSph and claimed to see extratidal members out to $3^\circ$ 
angular distance from the center of Draco in eastern direction. 
Piatek et al.\ (2001) investigated Draco by analysing color-magnitude 
diagrams of nine CCD fields at different positions and angular distances 
from its center. They found evidence for Draco stars in those fields which 
overlap with the standard limiting radius of $28\farcm3$ given by IH95 and 
reach somewhat beyond this limit. 
Furthermore, they argued that there might be weak evidence for Draco stars
also in their more distant fields at about 1.3 degrees from the center of 
Draco. \\ 

The Sloan Digital Sky Survey (SDSS; see York et al.\ 2000, Gunn et al.\ 1998,
Fukugita et al.\ 1996) has recently enabled us to investigate the Draco dSph 
through multi-color CCD photometry in a large contiguous field. Hereby, 
the distribution of its stellar content has been mapped more comprehensively 
and in more detail. As we show in this paper, the SDSS data do not confirm 
the conclusions from previous studies.\\

The paper is organized as follows:
\S~2 provides details of the SDSS data set and 
describes our method for the photometric selection of Draco 
members. In \S~3 we present surface density maps for two samples 
of member candidates and investigate them for signs of tail-like extensions.
In \S~4 we discuss fits of ellipsoidal models to the observed distribution 
of stars and determine the basic spatial parameters of the galaxy. 
\S~5 presents revised estimates of the total mass, luminosity and 
mass-to-light ratio.  In \S~6 we summarize the results and conclusions. \\

\begin{deluxetable}{ccl}
\tabletypesize{\footnotesize}
\tablecaption{Basic parameters of the Draco dsph galaxy \label{tab1}}
\tablewidth{0pt}
\tablehead{\colhead{Parameter} & \colhead{Value}   
& \colhead{Reference}}
\startdata
$\alpha,\delta$ (J2000)& $17^{\mathrm h}20^{\mathrm m}13\fs2$, $+57^{\circ}54'54''$ \\
$l,b$ & $86.4^\circ, +34.7^\circ$ & \\
$d$ & 71~kpc & Stetson (1979) + Carney et al.~(1992)\\
$v_r$ & $-$293.3~\kms\ & Armandroff et al.\ (1995) \\
$\sigma_{los}$ & 10.7~\kms\ & Armandroff et al.\ (1995) \\
$[$Fe/H$]$ & $-2.0$~dex & Lehnert et al.~(1992) \\
%$r_c$ & $7\farcm8\$ & this paper \\
%$r_t$ & $39\farcm2\$ & this paper \\
\enddata
\end{deluxetable}

\section{Photometric selection of Draco candidates}

SDSS is a multipurpose project which is designed to provide homogeneous, 
deep CCD imaging in five passbands ($u,g,r,i,z$) with contiguous coverage 
of $10^4$ square degrees in the Northern Galactic Cap.   
The SDSS data are thus ideally suited for studies of Galactic structure 
and local stellar systems (see, e.g., Yanny et al.\ 2000, 
Ivezic et al.\ 2000, Odenkirchen et al.\ 2001, Chen et al.\ 2001)
even though the primary goal of SDSS lies in extragalactic and cosmological 
research.\\

\subsection{The data set}

The field of Draco was observed in four neighboring great circle scans 
made with the SDSS CCD mosaic camera on April 3/4, and April 5/6, 2000 
(SDSS runs 1336, 1339, 1356, and 1359). 
The observations cover a four degree wide band across the Draco dSph 
galaxy and its surroundings and reach down to a limiting magnitude of 
$r^* \approx 23.0$. Our study focusses on the declination range of 
$55^\circ \le \delta \le 61^\circ$ and on objects recognized as point 
sources (based on their images being consistent with the local point 
spread function (PSF)).
The so-defined sample of about 290,000 objects has been investigated 
using PSF photometry from the standard SDSS data reduction pipeline 
(Lupton et al. 2001).
The typical photometric accuracies in the data set (median internal errors) 
are better than 0.03 mag in  $g$, $r$, and $i$ for sources brighter 
than 20.0~mag in these filter, and reach the level of 0.1~mag for source  
brightnesses of $u^*=21.0$, $g^*=22.3$, $r^*=22.0$, $i^*=21.5$, and 
$z^*=20.0$.
The data are preliminary in the sense that the calibration of the 
instrumental system onto to the nominal SDSS photometric system is still 
in progress. Therefore, we use asterisks to designate the currently used 
system of magnitudes (York et al.\ 2000). 
The differences that may occur with respect to a later final calibration 
will however be small (not larger than a few percent) and will thus not 
affect the results of this study in a significant way.\\    

The IR dust emission maps of Schlegel et al.\ (1998) show that some of 
the regions around the Draco dSph galaxy have non-negligible and spatially 
varying extinction. This is potentially dangerous for
our goal of detecting extended structures in the Draco population because 
it creates spatially varying shifts in the colors of the stars. 
The SDSS data include for each object and each filter an individual 
extinction estimate which is derived from the reddening proposed 
by the Schlegel et al.\ (1998) maps. 
Depending on the filter and the position of the object, the estimated 
extinction within our sample varies between 0.03 and 0.3 mag in 
$g$, $r$ and $i$, with local peaks of 0.4 in $u$. 
To remove reddening effects, the estimated extinction values have been 
subtracted from the measured magnitudes prior to any further analysis.  
We show the estimated extinction, $A_{g}$, in Fig.~1 as a function of 
position on the sky. For the main body of Draco, $A_{g}$ is, on average 
0.10~mag, while west of the galaxy and towards the southeast and the east 
there are regions where $A_{g}$ increases above 0.16 mag or 0.2 mag. 
The latter corresponds to a shift of about 0.05~mag in the color index 
$g-i$. \\

The crucial point in constructing a high-contrast map of a stellar system 
like Draco is efficient discrimination between the tracers of this system 
and the foreground field stars. Therefore, specific photometric selection 
criteria had to be worked out, using the central part of Draco as a training 
set.
In order to determine the photometric characteristics of the Draco population 
in the SDSS system we analysed the stars in an ellipse of semi-major axis 
$9'$ around the nominal center. These stars shown in Fig.~2. The members 
of the Draco dSph have colors in the range of $-0.5 \le g^*-r^* \le 1.0$ and  
$-0.5 \le r^*-i^* \le 0.5$ (see panels (a) and (b) of Fig.~2). 
The uncertainties of the $g^*$, $r^*$ and $i^*$ magnitudes of these stars
are approximately equal while the uncertainties of their $u^*$ and $z^*$ 
magnitudes are at least a factor of two higher. Thus the $g^*$,
$r^*$, and $i^*$ magnitudes were identified as the most suitable quantities 
for photometric filtering.  
The Draco members show a strong correlation between the color 
indices $g^*-r^*$ and $r^*-i^*$ which can be expressed in a simple linear 
relation between these indices.
This fact allowed us to facilitate the selection of Draco candidates 
by an orthogonal transformation from  $g^*-r^*$, $r^*-i^*$ to new 
indices $c_1$, $c_2$ as given in Eq.(1), similar to the indices used  
by Odenkirchen et al.\ (2001). 

\begin{mathletters}
\begin{eqnarray}
c_1&=& 0.927 (g^*-r^*) + 0.375 (r^*-i^*)\ \\
c_2&=& -0.375 (g^*-r^*) + 0.927 (r^*-i^*)\  
\end{eqnarray}
\end{mathletters}

Panels (c) and (d) of Fig.~2 present color-magnitude diagrams for the 
central part of Draco using these indices.   
The plane of $(c_1,i^*)$ shows the familiar pattern of an old stellar 
population, i.e. red giant branch, subgiant branch and horizontal branch. 
In the plane of $(c_2,i^*)$ the Draco stars simply scatter around $c_2 = 0$. 
This scatter is mainly due to photometric errors.  
An adequate selection in $c_2$ was thus obtained by deriving the median 
error $\overline{\sigma}_{c_2}$ of $c_2$ from the individual error estimates 
in $g^*$, $r^*$ and $i^*$ as a function of source brightness $i^*$ 
and by imposing the constraint $|c_2| \le 2\,\overline{\sigma}_{c_2}(i^*)$. 
The distributions of the color indices $u^*-g^*$ and $i^*-z^*$ were found
to be rather diffuse and turned out to be of little use for the selection 
of Draco candidates. Thus, for the reason of clarity, 
these indices were not included into the selection process.  \\

\subsection{The color-magnitude mask}
The most important means to discriminate Draco members from field 
stars is the combination of source brightness ($i^*$ in our case) and color 
index $c_1$. Our approach was to design an empirical filter mask in the 
color-magnitude plane of $(c_1,i^*)$ with optimal adaptation to the 
color-magnitude distributions of the Draco population and the field stars. 
Fig.~3 shows the density distribution $f_D$ of Draco stars in the plane 
of $(c_1,i^*)$ (Fig.~3a) and the corresponding distribution $f_F$ for 
field stars (Fig.~3b). 
These distributions were obtained through counts on a 
0.01\,mag $\times$ 0.05\,mag grid and subsequent weighted averaging within 
a radius of five grid points.
To get reliable count rates and a truly representative sample 
the distribution $f_D$ was derived from a relatively large area, i.e.\ 
an ellipse with semi-major axis $30'$ (ellipticity 0.3, position 
angle $90^\circ$) around the center of Draco.
The non-neglible, but well-defined, contribution from field stars in 
this area was subtracted by means of $f_F$. 
The distribution $f_F$ was obtained from stars with $\ge 2^\circ$ 
angular distance from the center of Draco. It exhibits two main 
concentrations:
The brighter one (max.\ density at $i^* \le 18, c_1\approx 0.4$) 
consists of slightly evolved thick disk stars and has little photometric 
overlap with the Draco population, the fainter one (max.\ density at 
$i^* > 19, c_1\approx 0.3$) consists of halo stars at and somewhat below 
the turn-off point of the subdwarf main-sequence. 
(see Chen et al.\ 2001).\footnote{ 
A third major field component would be M-dwarfs of the thin and thick disk. 
These do not appear in the diagram because they have $c_1 > 1.0$ and hence 
are of no concern.} 
We assume that the Draco population and the field population follow the 
distributions $f_D$ and $f_F$ in all parts of the field. 
This is presumably true for the field stars, but not necessarily for the 
Draco stars if there are population gradients. Nevertheless, one must adopt 
homogeneity as a working hypothesis.  
In Fig.~3c we show the population contrast $s = f_D/f_F$ as a function of 
position in the plane of $(c_1,i^*)$. 
The maxima of $s$ delineate those regions in the color-magnitude plane
with the highest fraction of Draco stars. To construct a map of Draco, 
both high contrast $s$ and large numbers of stars are necessary.
This suggests that a filter mask for the selection of Draco stars should 
include all points with $s \ge s_0$, with $s_0$ being a threshold that needs
to be optimized. 
Variations in $s_0$ define a one-parameter family of filter masks, which 
produce samples of different size and composition. 
Supplementary choices had to be made to fix the lower limit of $c_1$ and 
the upper limit of $i^*$, which are not well constrained by the distribution 
$s$. For the color index we chose
a cut-off at $c_1=0.0$~mag since there are not many Draco stars beyond this 
limit according to Fig.~2c. 
In $i^*$ we imposed a limit of $i^*\le 21.7$~mag, because the data from 
different observing runs have somewhat different limiting magnitudes and 
completeness limits that begin to show up by spatial inhomogeneities 
beyond $i^*=21.7$~mag.\\
   
To estimate the optimal threshold, $s_0$, we computed the signal-to-noise 
ratio (SNR) for the number of Draco stars from basic 
Poisson statistics of number counts with background subtraction:

\begin{eqnarray}
{\rm SNR(s_0)} = \frac{N(s_0) - w N_F(s_0)}{\sqrt{N(s_0) + w^2 N_F(s_0)}}
\end{eqnarray}

Here, $N$ is the total number of stars in the sample defined by $s_0$
and within a given region of Draco, $N_F$ the number of stars in the same 
sample, but in the region where the foreground field population is probed, 
and $w$ scales the angular extent of these two regions.  
Since the main goal was to determine the structure of the outer part 
of the Draco dSph, the SNR was computed and maximized for the annulus 
between $25'$ and $75'$ angular distance from the center of Draco. 
The maximum SNR was determined by scanning through a dense series of 
threshold values $s_0$.  
In Fig.~3c we show the color-magnitude masks in three typical cases: 
the mid-level contour corresponds to the mask that yields the highest SNR. 
The masks that correspond to the low-level and high-level contours 
of Fig.~3c yield larger and smaller samples, respectively, but both 
with lower SNR.  

\section{The spatial distribution of ``Draco-like'' stars}

We now discuss the distribution of Draco candidate stars in the plane of 
the sky by means of two special samples.  
The first sample (S1) has been obtained with the suboptimal filter given 
by the lowest contour of Fig.~3c; the second one (S2) has been selected with 
the optimal filter given by the mid-level contour of Fig.~3c. 
S1 has twice as many field stars and 1.4 times as many Draco stars as S2. 
The filters reduce the surface density of field stars from 
2.57~arcmin$^{-2}$ (unfiltered case) to 0.15~arcmin$^{-2}$ (S1) and 
0.074 arcmin$^{-2}$ (S2), i.e.\ by factors of 17 and 35, respectively.
In turn, the contrast between the Draco population and the residual field 
star population is strongly enhanced.  
Without any filtering the central stellar density contrast 
would be 5.7 (see Fig.~1), whereas for S1 and S2 it is 68.7 and 96.9,
respectively. Our samples thus reveal the spatial distribution of 
Draco with 15 to 20 times higher contrast than the previous map 
of IH95 which has a central density contrast of about 4.\\

We note that a few small areas (diameter $\le 1'$) in the central part of 
Draco show a pronounced lack of stars because the source detection and 
photometry has been hampered by the presence of a very bright foreground 
star. These regions have either been omitted from the analysis (e.g.\ in 
the fitting procedures) or else recovered by interpolation of the stellar 
density in the surroundings. 

\subsection{Surface density maps}
The spatial distribution of the stars in S1 and S2 is
shown in Fig.~4. Individual stellar positions are plotted as dots. 
Overlayed are contour lines of constant surface density. The surface density 
has been derived through counts on $3'$ by $3'$ grid and subsequent 
weighted averaging within a radius of two grid steps in the outer part and  
one grid step in the inner part of the field. 
The thin lines show contours at the level of 1\,$\sigma$ above the mean 
background density. Here, $\sigma$ is the rms of the background density 
fluctuations, indicating the lowest level at which Draco stars could become 
recognizable. This variance corresponds to surface densities of Draco stars 
of 0.035~arcmin$^{-2}$ for S1, and of 0.022~arcmin$^{-2}$ for S2, 
i.e. $3\times 10^{-3}$ of the central surface density of Draco.
Any significant detection of the Draco population requires at least 
2\,$\sigma$ above background, marked by the outermost thick contours in 
Fig.~4. 
Further contours are drawn at 5\,$\sigma$ and higher in 
order to reveal the shape of the galaxy at different radii. \\

Both samples show that the overall distribution of Draco stars is 
approximately ellipsoidal. With respect to the inner part the outer contours 
at $\le 5\sigma$ appear to be slightly deformed or shifted to the south-west
(see \S 4). Also, the contour lines for S2 are in general less smooth than 
those for S1. Nevertheless, there are no clear deviations from 
an elliptic shape down to the 2\,$\sigma$ contour. This contour reaches 
out to $30'$ angular distance from the center of Draco. \\

Strong departures from an elliptic shape are visible only at the 
1\,$\sigma$ level. Here, the map for the sample S2 shows more extended 
structures than the map for S1. There appear to be weak extensions in four 
directions out to radii of about $45'$. The ones pointing to the north-east
and south-west are filamentary while those in the south-east and the 
north-west are broader and may be described as a bar-like.
Beyond $45'$ distance from the center of Draco the field is sprinkled 
with numerous patches of 1\,$\sigma$ overdensities. About one third of 
them are visible also on the 2\,$\sigma$ level. These patches are 
distributed rather uniformly, and do not form large connected structures. 
Altogether they cover an area of no more than 30\% of the total field.
This suggests that these patches simply reflect random fluctuations 
of the residual field star population. 
A closer look at the statistics of the outer fields supports this view. 
In Fig.~5 we show histograms and cumulative distributions of the local 
S2 field star densities measured in non-overlapping grid cells of 
$6' \times 6'$ and $9' \times 9'$ beyond $45'$ distance from the 
center of Draco. The open circles show Poisson distributions for 
the observed mean values. 
The measured fluctuations are in close agreement and indeed statistically 
consistent with the corresponding Poisson models. 
A Kolmogorov-Smirnow test shows that the opposite hypothesis, 
namely that the empirical distribution is different from 
the Poisson model, has in both cases a probability of not more than 2\%. 
Thus we conclude that the number of grid cells with surface densities 
of e.g.\ 1\,$\sigma$ or 2\,$\sigma$ above the mean does not significantly 
exceed the number expected from Poisson noise of a random distribution. 
(cf.\ Fig.~5b). Hence there is no evidence, and indeed little room, for 
overdensities that can be ascribed to Draco.  
Another way to show that the density enhancements around 
Draco in the maps of Fig.~4 are random fluctuations is by comparing the 
observations with a random number generated star field of equal mean density.
Fig.~6 shows such a simulated field corresponding to sample S2. The surface 
density has been derived using the same pixel grid and smoothing and the 
density contours are drawn on the same levels as for S2 in Fig.~4b.     
The $1\sigma$ and $2\sigma$ contours in the simulated map show structures
that are very similar in size and number to those in the maps drawn from 
the observational data. \\
 
Density fluctuations of the field star population must of course  
also occur in the region where the Draco galaxy is seen. 
The superposition of these fluctuations with the low density of Draco 
stars in the outermost parts of the galaxy is likely to create distortions 
in the 1\,$\sigma$ contour around Draco. This is also nicely revealed by
the simulated example of Fig.~6.     
Since the four weak extensions which are visible in Fig.~4b and which have 
been described above are not 
substantially larger than the sizes of the 1\,$\sigma$ patches in the outer 
field it is very likely that these extensions are also due to fluctuations 
in the density of residual field stars.  
At least, these structures are no significant indication of the presence 
of Draco stars.

In summary, we find that down to the level of 2\,$\sigma$ above the 
background density of our maps the Draco dSph has a fairly regular 
shape, and that there are no detectable traces of tail-like extensions down 
to $3\times 10^{-3}$ of the central density. \\

\subsection{Mean surface densities along major and minor axis}
The stringent non-detection of tidal tail features on our new Draco maps 
contradicts the previous conclusions of SKH97, who derived mean surface 
densities in fields of $23' \times 23'$ along the principal axes of Draco. 
For a direct comparison with the SKH97 results, we determined mean surface 
densities from our data in fields of the same size and 
position along the same axes (major axis at position angle $82^\circ$).
These are plotted in Fig.~7 as a function of position 
along the major and minor axes with respect to the center of the galaxy. 
As in SKH97, the mean density in each field is normalized to the 
mean density in the central field, and an estimate of the foreground 
density is subtracted. Our estimate of the mean foreground density,
however, relies not only on one remote field as in SKH97, but on all stars 
with more than $2^\circ$ distance from the center of Draco. 
The fields at $20'$ mean angular distance from the center of Draco have 
scaled densities of about 0.15 with respect to the mean density 
of the central field, compared to 0.12 reported by SKH97. 
In contrast to SKH97, we do not observe significant overdensities along 
the major axis far from the center of Draco in either of our two samples. 
In the more distant fields, the mean density agrees with the density level 
of foreground stars within the statistical errors. There is no general 
difference between the mean densities measured along the major axis 
and those measured along the minor axis beyond a distance of $40'$ from 
the center. However, there are of course statistical fluctuations 
in the mean densities of the individual fields along both axes. 
Fig.~7 shows that the reference foreground field for SKH97
(i.e., at $1^\circ$ north of Draco) is the one with the lowest mean 
density in our series of fields. We would obtain significant 
overdensities in most of the other fields if we chose this particular 
field as a reference. This suggests that the overdensities reported by 
SKH97 reflect to a large extent an underestimated foreground density. 
The resolution of this apparent discrepancy shows once more the power of
wide area data.

\section{Size and structure of the Draco dSph galaxy}
To quantify the size, shape and orientation of the Draco dSph we fit 
two-dimensional models to the observed surface densities. 
Herein we use the empirical King profile (King 1962, hererafter K62; 
Eq.\,3a), the theoretical King profile (King 1966, hereafter K66), and 
a generalized exponential (Sersic 1968; Eq.\,3b): 

\begin{mathletters}
\begin{eqnarray}
\Sigma(r) = \Sigma_1 \cdot \left(\frac{1}{\sqrt{1+(r/r_c)^2}} - 
\frac{1}{\sqrt{1+(r_t/r_c)^2}} \right)^2 
\end{eqnarray}
\begin{eqnarray}
\Sigma(r) = \Sigma_0 \exp \left( -(r/r_0)^n \right)
\end{eqnarray}
\end{mathletters}

In all cases $r=\sqrt{x^2+y^2/(1-e)^2}$ is the elliptical radius and $x$ 
the coordinate aligned with the major axis.
The profile Eq.\,3a has a finite limiting radius $r_t$ which is meant to 
describe tidal truncation. It is often used in simple models of globular 
clusters. 
The theoretical models of K66 have projected density profiles that resemble 
those of Eq.\,3a, but with a somewhat shallower decline towards the cut-off 
radius. These profiles cannot be expressed explicitly and were generated by
numerical integration following the description by Binney \& Tremaine 
(1987, p.232).  
The profile Eq.\,3b is of infinite extent and has been found to fit 
elliptical galaxies in a wide range of absolute magnitudes 
(Young \& Currie 1994, Jerjen \& Binggeli 1997).      
The three types of models were fit to the surface density distributions 
of S1 and S2 by weighted least squares. The surface densities were sampled 
on a $3'\times 3'$ grid of non-overlapping cells. A few cells which fell 
on positions that are blended by bright stars were excluded from the fit.
Weights were derived from Poisson noise, with the expected number of stars 
given by the fitted model. 
The weights were iteratively adjusted to yield a consistent solution. 
The best-fit parameters are given in Table~2 and the best fits are 
shown in Figs.~8 and 9.\\

The values of $\chi^2/(N-f)$ in Table~2 
\footnote{$(N-f)$ denotes the degrees of freedom of the fit and is 
$\approx 4350$.} as well as the residuals plotted in 
Figs.~8b,8d and the profiles in Fig.~9 demonstrate that the three types 
of models all provide good fits.
Moreover, the model parameters obtained from the two different data 
samples are consistent within their formal errors. 
Hence the results do not depend critically on the particular choice of the 
color-magnitude filter. 
The best-fit ellipticity is $0.29\pm0.03$, confirming earlier results of 
IH95 and Hodge (1964). 
The major axis position angle is $88^\circ (\pm 3)$ (J2000.0), in slight
disagreement with the earlier determination of IH95 (see Table 2).\\

The most interesting parameters are those describing Draco's 
size. Here our results differ substantially from those given by IH95.
The fit of the K62 model yields for both samples a core radius that 
is about 15\% smaller and, more importantly, a limiting, or tidal, radius 
that is about 40\% larger than the previous standard value of IH95 
(see Table 2).  
The latter has important consequences: (1) It shows that Draco is more 
extended and therefore also more massive than previously assumed (see \S 5). 
(2) It shows that some of the previous detections of Draco stars beyond 
the former standard radius of 28\farcm3 are in fact not extratidal if we 
define the (actually unknown) tidal radius of the galaxy as usual by the 
limiting radius of the best-fit King profile.

To demonstrate the clear evidence for a larger spatial extent of Draco 
in our data, we show in Fig.~9 the observed radial density profiles for 
S1 and S2. The densities are tabulated in Table 3.
The profiles were derived from star counts in elliptical annuli. The 
logarithm of the mean density above background is plotted versus the 
logarithm of the rms of the outer and inner radius of each annulus.  
The radii refer to the major axis.
The mean background density was obtained as one of the parameters in 
the least-squares fitting of the two-dimensional model distributions 
where it is determined by the surface density data in the outer parts of 
the field.
Error bars at each bin indicate the statistical count rate uncertainty 
(quadratically combined uncertainties of $\Sigma$ and $\Sigma_{bg}$)
and show that all bins except the outermost one have significant 
densities of Draco stars. Note that the counts in these bins trace the 
profile of the galaxy over three orders of magnitude and thus reflect a 
ten-fold increase in the dynamic range compared to IH95. 
Three relevant model curves are overlayed as lines in Figs.~9a and 9b: two 
King models (dashed lines) and one generalized exponential model 
(solid lines). 
Clearly, the K62 model of IH95 (short dashed lines) fails to describe our 
data, as there are four consecutive radial bins with significantly higher 
surface density. 
The new best-fit K62 models (long dashed lines) represent the observed 
densities (within their statistical uncertainties) down to the outermost 
bin.
There are no hints for the existence of a break in the slope of the 
radial profile.  
The models with the K66 surface density profile fit the data with the 
same overall quality but provide a better approximation to the 
outermost significant bin ($34' < r \le 40'$) since the decline of the 
density profile is less steep. Extra\-polation with the best-fit K66 profile 
thus yields a limiting radius of about $50'$ instead of about $40'$ from 
K62.     
The best-fit generalized exponential model is characterized by a radial 
scale length of $r_0 = 7\farcm3$ (S1) or $7\farcm6$ (S2) and an 
exponent $n= 1.2$. This profile fits the outermost significant 
bin in the same way as the best-fit K66 models. For sample S2 the fit 
of the generalized exponential even attains a slightly better 
overall $\chi^2$. 
This reveals that there is no compelling evidence for a real cut-off in 
the radial profile of the Draco dSph within the current observational 
limits. In other words, despite the improved contrast of our density maps 
a distinct edge of the galaxy is possibly still undetected. 
Our result of $r_t \simeq 40'$ from the best-fit K62 model may thus 
underestimate Draco's real radius but it puts at least a firm lower limit 
on any possibly existing true density cut-off.

\begin{deluxetable}{clccccccccc}
\tabletypesize{\footnotesize}
\tablecaption{Parameter values of best-fit ellipsoidal models of Draco 
\label{tab2}}
\tablewidth{0pt}
\tablehead{\colhead{Sample}&\colhead{Model} & \colhead{$\alpha_c$} & 
\colhead{$\delta_c$} &\colhead{e}&\colhead{PA}&
\colhead{$r_c$}&\colhead{$r_t$}&\colhead{$r_0$}&\colhead{$n$}&
\colhead{$\chi^2/(N-f)$}}
\startdata
S1 & King62 &  260\fdg055 & +57\fdg917 & 0.28 & 89$^\circ$ & 7\farcm4 & 39\farcm6 &- & - & 1.019 \\
  &  &$\pm 0.006$&$\pm 0.003$&$\pm$0.01 &$\pm$3 &$\pm$0.2 &$\pm$0.8 &- & - & - \\
S1 & King66 &  260\fdg055 & +57\fdg917 & 0.28 & 89$^\circ$ & 8\farcm3 & 49\farcm6 &- & - & 1.018 \\
  &  & - & - & - & - &$\pm$0.2 &$\pm$1.3 & - & - & - \\
S1 & Sersic &  260\fdg055& +57\fdg916 & 0.28 & 89$^\circ$ & - & - &     7\farcm3 & 1.2 & 1.017 \\
  &  &$\pm 0.006$&$\pm 0.003$&$\pm$0.01 &$\pm$3 & - & - &$\pm$0.1 & $\pm$0.1 & - \\
\\
S2 & King62 &  260\fdg055 & +57\fdg915 & 0.30 & 88$^\circ$ & 7\farcm7 & 40\farcm1 & - & - & 0.995 \\
  &   &$\pm 0.008$&$\pm 0.003$&$\pm$0.02 &$\pm$3 &$\pm$0.2 &$\pm$0.9 & - & - & -\\
S2 & King66 &  260\fdg055 & +57\fdg915 & 0.30 & 88$^\circ$ & 8\farcm7 & 49\farcm4 & - & - & 0.994 \\
  &   & - & - & - & - &$\pm$0.3 &$\pm$1.4 & - & - & -\\
S2 & Sersic &  260\fdg055 & +57\fdg915 & 0.30 & 88$^\circ$ & - & - &     7\farcm6 & 1.2 & 0.988 \\
  &   &$\pm 0.008$&$\pm 0.003$&$\pm$0.02 &$\pm$3 & - & - &$\pm$0.1 & $\pm$0.1 & -\\
\\
\\
IH95 &King62& - & - & 0.29 & 82$^\circ$ & 9\farcm0 & 28\farcm3 & - & - & -\\
     &      & - & - &$\pm$0.01&$\pm$1&$\pm$0.7&$\pm$2.4& - & - & -\\
\enddata

Note that the meaning of $r_c$ for K62 and K66 is different and depends on 
the concentration of the model.
\end{deluxetable}

\begin{deluxetable}{ccccccc}
\tabletypesize{\footnotesize}
\tablecaption{Radial surface density profiles
\label{tab3}}
\tablewidth{0pt}\
\tablehead{\colhead{}&\colhead{}&\colhead{}&\multicolumn{2}{c}{S1}&
\multicolumn{2}{c}{S2} \\
\colhead{$r_{in}$}&\colhead{$r_{out}$} & \colhead{$r_m$} & 
\colhead{$\Sigma$} &\colhead{$\sigma(\Sigma)$}&\colhead{$\Sigma$}&
\colhead{$\sigma(\Sigma)$}\\
\multicolumn{3}{c}{arcmin}& \multicolumn{2}{c}{arcmin$^{-2}$} &
\multicolumn{2}{c}{arcmin$^{-2}$}}
\startdata
     0.0  &  1.0 &  0.71 & 10.250 &  2.137 &  7.181 &  1.795 \\
     1.0  &  2.0 &  1.58 &  8.467 &  1.121 &  5.087 &  0.872 \\
     2.0  &  3.0 &  2.55 &  7.219 &  0.802 &  4.937 &  0.666 \\
     3.0  &  4.0 &  3.54 &  7.448 &  0.689 &  4.937 &  0.563 \\
     4.0  &  5.0 &  4.53 &  7.130 &  0.594 &  4.388 &  0.468 \\
     5.0  &  6.0 &  5.52 &  4.740 &  0.438 &  3.142 &  0.358 \\
     6.0  &  7.0 &  6.52 &  5.142 &  0.420 &  3.452 &  0.345 \\
     7.0  &  8.0 &  7.52 &  3.654 &  0.329 &  2.723 &  0.285 \\
     8.0  & 10.0 &  9.06 &  3.181 &  0.198 &  2.132 &  0.163 \\
    10.0  & 12.0 & 11.05 &  2.380 &  0.155 &  1.642 &  0.129 \\
    12.0  & 14.0 & 13.04 &  1.363 &  0.108 &  0.941 &  0.090 \\
    14.0  & 16.0 & 15.03 &  1.099 &  0.090 &  0.778 &  0.076 \\
    16.0  & 18.0 & 17.03 &  0.629 &  0.064 &  0.442 &  0.054 \\
    18.0  & 22.0 & 20.10 &  0.485 &  0.037 &  0.278 &  0.028 \\
    22.0  & 28.0 & 25.18 &  0.300 &  0.021 &  0.180 &  0.016 \\
    28.0  & 34.0 & 31.14 &  0.193 &  0.015 &  0.110 &  0.012 \\
    34.0  & 40.0 & 37.12 &  0.165 &  0.013 &  0.088 &  0.009 \\
    40.0  & 60.0 & 50.99 &  0.150 &  0.006 &  0.078 &  0.004 \\
\enddata

Notes: $r_m = \sqrt{0.5 (r_{in}^2 + r_{out}^2)}$\\
The ``background'' surface densities are: \\
$\Sigma_{bg} = 0.1511\pm 0.0020$~arcmin$^{-2}$ (S1) \\
$\Sigma_{bg} = 0.0760\pm 0.0014$~arcmin$^{-2}$ (S2) \\
\end{deluxetable}

The model fits provide a useful reference to 
identify peculiarities in the shape of the observed distribution of stars.
In Figs.~8a and 8c we compare the contours of the observed density 
distribution with the elliptic contours of the best-fit exponential model.
Apart from the filamentary and barlike extensions on the $1\,\sigma$ level 
(see \S\,3) one recognizes a less extended box-like deformation 
of the galaxy with respect to the model in the south-western quadrant. 
This feature shows up in both samples and is present also in the $2\,\sigma$ 
and $3\,\sigma$ contours. 
The residual maps (Figs.~8b and 8d) trace these deformations to two patches 
of enhanced surface density which are located about 
$25'$ south and southwest of the center of Draco. 
Their $2 \sigma$ significance might indicate real irregularities in the 
stellar distribution of Draco.   
In order to quantify the amount of asymmetry we counted the number 
of stars in two opposite sectors separated by a straight line through the 
center of the galaxy (bisector method). From the (background-subtracted) 
numbers $n_1,n_2$ in the two sectors one can define an asymmetry parameter 
$A$ as:

\begin{eqnarray}
A = \left|\frac{n_1-n_2}{n_1+n_2}\right|
\end{eqnarray}

For $A<<1$, the uncertainty of $A$ is $\sigma_A = 1/\sqrt{n_1+n_2}$. 
In the inner part of Draco ($r < 15'$) both S1 and S2 yield no significant 
asymmetry ($A < \sigma_A$, $\sigma_A \sim 2.5\%$) for all position angles 
of the bisector. 
For $15' < r < 40'$ we obtain maximum values of 
$A = (6.1 \pm 3.4)\%$  for sample S1 and $A = (9.0 \pm 4.6)\%$  
for sample S2, both with bisector position angles of about $150^\circ$.
The outer part of Draco thus seems to have an asymmetric stellar distribution 
at the 10\% level ($2 \sigma$ significance). Note that any tidal deformation 
is expected to be symmetrical with respect to the center.

\section{Mass, luminosity and global mass-to-light ratio} 

\subsection{Estimating the total mass}
The new values for the spatial parameters $r_c, r_t$ of Draco 
lead to a revised estimate of the total mass $M$ of the Draco dSph. 
We estimate $M$ in the usual way by means of the mass of an equivalent
single-component King sphere. Following K66, this mass is given 
by:\\ 

\begin{eqnarray}
M = \frac{9}{4{\pi}G}\ r_c\ \mu\ (\beta\,\sigma_{los})^2
\end{eqnarray}

Herein $\mu$ denotes the normalized mass (see Eq.\,38 of K66) which is 
only a function of the concentration of the system, $r_c$ the linear core 
radius, $\sigma_{los}$ the observed line-of-sight velocity dispersion, 
and $\beta$ a correction factor to link the observed dispersion to King's 
parameter $\sigma_0$. 
We recall that this way of estimating the mass of a stellar system 
relies on three basic assumptions: (1) dynamical equilibrium, 
(2) proportionality of mass and light distribution, and (3) the 
adequacy of a spherically symmetric and isotropic model in 
general and a King model in particular. 

Our values for $r_c$ and $r_t$ from the best-fit K66 profile imply a 
concentration of $c = \log(r_t/r_c) = 0.77\pm0.03$. The normalized mass 
thus is $\mu=6.8\pm0.5$ (as opposed to $\mu = 2.6$ used in IH95). 
The core radius of the dynamical model is appropriately chosen as the 
geometric mean $r_c = r_{c,obs}\sqrt{1-e}$ since this preserves the 
projected area of the system when replacing the observed ellipsoidal 
system by the spherical model. 
To determine the relative change in the mass $M$ that is due to changes in 
$r_c$ and $r_t$ it suffices to evaluate the product $r_c\cdot\mu$, using 
$r_c$ in angular measure. We thus find that $M$ increases by a factor of 
2.5 when using the revised values of $r_c$ and $r_t$ instead of the former 
ones. If the rest of Eq.\,5 is to be kept unchanged, former estimates of 
$M$ could be updated by just multiplying with this factor. However, we 
consider it worthwhile to assess also the values of the other parameters 
involved in Eq.\,5.  \\

The physical value of $r_c$ requires an estimate of Draco's heliocentric 
distance $d$. 
Adopting the apparent brightness of the red horizontal branch measured 
by Stetson (1979) ($V_{\rm RHB} = 20.07 \pm 0.03$, $E(B-V)=0.03$), the 
empirical HB luminosity relation of Carney, Storm \& Jones (1992) 
($M_V = 0.15\,[{\rm Fe/H}] + 1.01$) and a mean metal abundance of 
$[{\rm Fe/H}]=-2.0$ (see the compilation by Lehnert et al. 1992) 
one obtains $(m-M)_0 = 19.26\pm 0.20$ and hence $d = 71\pm 7$~kpc.\footnote{
The uncertainty of 0.2 mag in the distance modulus accounts for 
discrepancies between horizontal branch luminosity relations derived   
by different authors and methods.} 
This yields a linear core radius of $r_c = 148\pm 15$~pc (geometric mean). 
From spectroscopic measurements of 91 Draco members, the most extensive 
data set of radial velocities in Draco that is currently available, 
Armandroff et al.\ (1995) derived line-of-sight velocity dispersions of 
$10.7\pm 0.9$~\kms (complete sample) and $(8.5\pm 0.7)$~\kms (by omission 
of three extreme velocities). Olszewski et al.\ (1996) showed that these 
results are robust against contamination by orbital motion of binaries. 
Most of the stars of this radial velocity sample are located within an 
ellipse with semi-major axis of $1.5 r_c$. 
By Monte Carlo simulation of a K66 dynamical model with $c=0.77$ we find 
that a correction factor $\beta \simeq 1.35$ applies if the line-of-sight 
velocity dispersion is measured with stars that cover the range 
$r \le 1.5\,r_c$. Inserting all these values into Eq.\,5 we obtain 
estimates of the total mass of the Draco dSph of 
$3.5\pm0.7 \times 10^7 M_\odot$ and $2.2\pm0.5 \times 10^7 M_\odot$ 
depending on which of the two velocity dispersions are used. \\  

As expected, our result of $3.5\times 10^7 M_\odot$ exceeds some 
of the previous estimates derived on the basis of the parameters of IH95,
but the increase is less than by the factor of 2.5 mentioned above.  
Hargreaves et al. (1996), for example, obtained $M = 2.6\times 10^7 M_\odot$
with the same method of King model mass estimation, using $r_c$ and 
$r_t$ from IH95 and a velocity dispersion very near to the 10.7~\kms of 
Armandroff et al., but with a different choice for $\beta$. 
Mateo (1998) and van den Bergh (2000) give the total mass of Draco as 
$2.2\times 10^7 M_\odot$ referring to similar values for the spatial and 
kinematic parameters, but without presenting their calculations in detail. 
The mass estimate of $4.4\times 10^7 M_\odot$ given in IH95 is 
substantially higher than ours and those by other authors. 
This seems confusing but can be traced to their adopted dispersion 
of 13.2~\kms, which has been invalidated by the measurements of 
Armandroff et al.\ and Hargreaves et al.
Our revised estimates of Draco's mass benefit from an improved knowledge 
of the spatial as well as of the kinematic parameters and hence are more 
precise. Nevertheless, their validity remains limited to the point that 
the underlying basic assumptions (see second paragraph of \S~5.1) 
must hold, at least approximately.\\

The assumption that ``mass follows light'' might be wrong in the sense 
that the dark matter component which is likely to exist and to dominate
the matter content of Draco (see \S 5.2 and \S 6.2) could have a less 
concentrated distribution than the luminous matter, as is often found to 
be the case in stellar systems that contain dark matter.  Thus it is 
interesting to see how the estimate of the total mass changes under such 
circumstances. In the most extreme case the dark matter component would 
have a constant density $\rho_{DM}$ (out to radius $r_t$) and the stars 
would move as test particles in the harmonic potential of this homogeneous 
dark matter distribution. Here it is convenient to approximate the space 
density of the stars in the core region of Draco by a Gaussian radial 
density  profile, i.e., $\rho = \rho_0 \exp(-(r/r_0)^2/2)$, with scale 
length $r_0 = r_c/\gamma$ and $\gamma \simeq 1.42$ for $c= 0.77$.
From the Jeans equation (see Binney \& Tremaine 1987, p.204) it follows 
that a system with Gaussian density profile in a harmonic potential has a 
velocity distribution with constant dispersion 
$\sigma = \sqrt{4\pi G \rho_{DM}/3}\ r_0$ (see als0 Pryor \& Kormendy 1990). 
Setting $\sigma = \sigma_{los}$, the mass $\widetilde{M}$ of the dark 
component out to radius $r_t$ and the ratio between this $\widetilde{M}$ 
and the King mass $M$ of Eq.5 are

\begin{eqnarray}
\widetilde{M} = \frac{1}{G}\ \frac{r_t^3}{r_0^2}\ \sigma_{los}^2
\end{eqnarray}

\begin{eqnarray}
\frac{\widetilde{M}}{M} = 
\frac{4\pi}{9}\ \left(\frac{\gamma}{\beta}\right)^2\ 
\left(\frac{r_t}{r_c}\right)^3\ \frac{1}{\mu}
\end{eqnarray}

Putting the appropriate parameter values into Eq.\,7 one finds 
$\widetilde{M}/M \simeq 45$.
This example demonstrates that the mass estimate derived by Eq.\,5, 
assuming that all matter in Draco follows the distribution of the stars, 
should be regarded as a lower limit because a more shallow distribution 
of the dark matter component would result in a considerably higher total 
mass. 

\subsection{Luminosity and mass-to-light ratio}

We determined the total luminosity of the Draco dSph by (1) summing up 
the $i$ band fluxes of all possible Draco members in the magnitude 
interval $16.0 \le i^* \le 21.5$, (2) subtracting the 
expected contribution of flux from foreground stars, and (3) 
adding an estimate of the missing flux from Draco stars at $i^* > 21.5$.   
The flux integral was calculated in the region of the color-magnitude 
plane that is outlined by the solid line in Fig.~10, and for 
major axis radii varying from $10'$ to $50'$. The foreground contribution 
was estimated by an analogous integration in a ring around Draco extending
from  $1^\circ$ to $2^\circ$. 
The foreground-corrected luminosity function of Draco is shown in Fig.~10b.
After foreground subtraction the integrated magnitude for Draco 
converges to $m_{i} = 11.09 \pm 0.03$ at a radius of $r = 20'$. 
At $r > 20'$ the integrated magnitude deviates by not more than 
$\pm 0.03$~mag from this value. 
The uncertainty of $m_{i}$ was estimated by splitting the fields into 
four quadrants and calculating the flux integral for each quadrant 
separately. The resulting rms was 15 \%, implying an uncertainty of
0.08~mag for the determination of $m_{i}$ from the entire field. 
As a further check we mention that an erroneous inclusion or omission of 
up to three stars at the bright end of Draco's giant branch ($i^* = 16$) 
would change $m_{i}$ by not more than 0.04~mag, which is only half of
the quoted error bar.\\  

The missing flux from faint stars beyond our magnitude cut-off was estimated 
by analysing deep HST observations of a small field near the center of 
Draco (Grillmair et al.\ 1998). We retrieved these observations from 
the HST archive and reduced them with the software package HSTphot 
(Dolphin 2000). The resulting photometry covers the
magnitude range $18.5 \le I \le 26.0$ and has an overlap of 2.8 mag 
with the SDSS data in the range $ 18.5 \le I \le 21.3$ or
$18.9 \le i^* \le 21.7$. The luminosity function derived from the 
HST data is shown by the dashed line in Fig.~10b.  
We integrated the $I$ band counts from HST  
between $I=21.3$ and $I=26.0$ and extrapolated the integral to 
$I=27.75$ \footnote{Deep HST observations of globular clusters
by Piotto et al.\ (1997) show that the maximum of the luminosity function 
normally lies at about $M_I = 8.5$, i.e.\ $I=27.75$ for Draco.} assuming a 
constant slope of the luminosity function. 
We then scaled the result with the ratio between the number of Draco 
stars in the (global) SDSS sample and the number in the (local) HST sample, 
both counted in the overlap interval. 
This yields an estimate for the missing flux in the SDSS data of
0.59 ($\pm 0.08$) mag. Correcting for the faint stars in Draco thus  
leads to $m_{i} = 10.49\pm 0.10$ for the total magnitude. 
By subtraction of the distance modulus ($(m-M)_0 = 19.26$, see \S~5.1) 
$m_i$ translates into a total absolute magnitude of $M_{i}=-8.77\pm 0.20$.
The uncertainty of the absolute magnitude is dominated by the uncertainty 
of the distance modulus of 0.2~mag. 

For conversion to solar units we used basic photometric data for the Sun 
given in Landolt-B\"ornstein (1981) and an empirical transformation relation 
between standard photometry and SDSS photometry (Grebel et al.\ 2001):

\begin{eqnarray}
i^* = 0.33(\pm 0.02) + I_C + 0.19(\pm 0.02)\,(V - I_C)
\end{eqnarray}

This relation was derived by comparing SDSS data with $VI_C$
photometry in the photometric standard fields of Stetson (2000).
For the Sun we have $M_{V,\odot} = 4.87$ and a color index 
$(V-I)_\odot = 0.81$ in the Johnson system, which,
according to the transformation formulae of Fernie (1983), 
corresponds to $(V - I_C)_\odot = 0.64$ in the Johnson-Cousins system.  
Putting this into Eq.\,8 one obtains $M_{i^*,\odot} = 4.68$. 
Our result for the luminosity of Draco
thus reads $(L/L_\odot)_{i} = (2.4 \pm 0.5) \times 10^5$. 
This is in good agreement with earlier determinations of $(L/L_\odot$) 
made in a different wavelength range, e.g. by IH95 who give the  
luminosity of Draco as $(L/L_\odot)_V = (1.8 \pm 0.8) \times 10^5$. 
We emphasize that the determination of the luminosity does not depend on 
critical assumptions and models and is in this sense 
more general than the determination of the total mass. \\

Combining the luminosity and the mass estimates of \S~5.1
we find total mass-to-light ratios $M/L_{i}$ of either $146\pm 42$ 
or $92\pm 28$ in solar units. For comparison, the values of $M/L_V$ 
given by Armandroff et al.\ (1995), were $90$ and $57$ in solar units
(with respect to the $V$ band). 
As for the values for the total mass, the mass-to-light ratios are 
valid only on the proviso that the assumptions mentioned in \S~5.1 
hold. Note however, that the true mass of the system would need to be 
smaller by a factor of 30 or more in order to achieve a normal mass-to-light 
ratio of 1 -- 3 in solar units. On the other hand it has been shown that 
a distribution of dark matter that is less concentrated than the 
distribution of the stars would result in an increase of the total 
mass by up to a factor of 45 and thus lead to a further increase of the 
mass-to-light ratio.  
\\

An additional argument for a high mass-to-light ratio comes from
the size limit that is imposed by the galactic tidal forces. 
This argument is independent from the above direct estimate 
of Draco's total mass since it relies only on the observed maximum 
spatial extent and the luminosity. 
For a satellite of mass $M$ orbiting in a spherically symmetric 
potential at (constant) galactocentric distance $R$ there exists a 
well-defined theoretical size limit that is given by the Lagrange 
radius $r_L$ characterizing the force balance between internal gravity, 
differential centrifugal forces and the galactic tidal field. 
Taking a logarithmic potential with circular velocity $v_c = 220$~\kms
as a standard model for the gravitational potential in the Galactic halo, 
the Lagrange radius is given by the relation:       

\begin{eqnarray}
r_L^3 = \frac{GM}{2 v_c^2}\cdot R^2 
\end{eqnarray}

Fig.~11 shows $r_L$ as a function of galactocentric distance $R$
for different satellite masses $M$ (dashed lines) and different 
mass-to-light ratios (solid lines), adopting the observed luminosity 
$(L/L_\odot)_{i} = 2.4 \times 10^5$ of Draco. 
The fact that no signs of tidal perturbation have been detected
suggests that at least the currently observed part of Draco's stellar 
distribution does not reach beyond the Lagrange radius. 
Therefore we have the condition $r_t \le r_L$. 
The measured limiting radius $r_t$ is drawn as the hatched bar in 
Fig.~11. The galactocentric radial velocity of Draco
\footnote{Note that the line of sight from the Sun to 
Draco is close to Draco's galactocentric radius vector},
which is near $-$100~\kms, indicates that Draco is not on a circular orbit. 
Nevertheless, one can use $r_L$ for obtaining an approximate size limit. 
The appropriate distance $R$ at which to take $r_L$ is in this case however 
not evident; it might be the pericentric distance $R_p$ rather than the 
current distance $R_0$. 
Since $R_p$ cannot be estimated without reliable information on the 
tangential velocity, one has to consider a wide range of distances 
$R \le R_0$. Fig.~11 reveals that for any such choice of $R$ a normal 
mass-to-light ratio of $M/L\le 3$ yields Lagrange radii that are 
substantially smaller than the observed limiting radius $r_t$ for which 
we here take the conservative value of $40'$ or 820\,pc.
Hence a normal value of $M/L$ is ruled out within the context of our 
assumptions. 
Possible values of $M/L$ in the sense that $r_t < r_L$ holds for 
distances well below $R_0$ lie near or above $M/L=30$. 
With $M/L \approx 100$, i.e. a satellite mass of about 
$3\times 10^7~M_\odot$ as derived by direct mass estimation with 
Eq.~5, the observed size of Draco remains below the theoretical limit 
of $r_L$ down to galactocentric distances of 20 to 25~kpc. 
This means that a mass-to-light ratio of about 100 or higher can 
prevent tidal stripping of stars from the outermost part of Draco 
even in relatively close approaches to the Milky Way.  
It must however be noted that this is a stationary picture. In reality 
a passage of the Galactic center on an elongated orbit could have the 
additional effect of heating the stellar population 
of the satellite due to the non-stationarity of the tidal forces. 
This kind of influence on the internal kinematics of the satellite 
system is not accounted for by the simple Lagrange radius criterion.

\section{Discussion and conclusions}

\subsection{The missing extratidal component}

From the results of previous studies, in particular those of IH95
and SKH97, one might have expected that a high 
contrast map of the Draco galaxy down to magnitude 22 would 
show clear signs of tidal features.  
Our analysis of the SDSS data shows that this is not the case. 
Structures which could be interpreted as indicating tail-like extensions 
appear only on the 1\,$\sigma$ level of our most sensitive map, i.e., at
$2.8\times 10^{-3}$ of the galaxy's central density. Since these structures 
go into four different directions their interpretation in terms of tidal 
tails is problematic. The more bar-like structures seen towards 
the south-east and the north-west would appear as the preferred candidates
for such tails. However, by statistical analysis of the density variations 
in the surrounding field we find that the observed amount of structure at
1\,$\sigma$ above background density is easily explicable by normal Poisson 
noise. Therefore, none of the extensions in the vicinity of the main body 
of Draco nor any of the overdense patches seen at larger angular distance 
are likely to be an external part of its stellar population.
    
On the other hand, the radial profile of the surface density shows that 
the ordinary spatial extent of the Draco galaxy is substantially larger 
than was revealed by previous measurements with lower contrast. Going down 
to about $10^{-3}$ of the central density, the observed radial density 
profiles directly reveal that Draco extends to about $40'$. 
The same is found by fitting ellipsoidal models to the observed surface 
density distribution. 
A K62 profile yields a best-fit $r_t$ of $39\farcm6$ (S1) and $40\farcm1$ 
(S2), which is 1.4 times the previous determination (all radii referring to 
the major axis). Using the profile of the K66 theoretical models one obtains 
an even larger estimate of $r_t =49\farcm5$.  
The observations do however not provide particular support for a 
truncated radial profile.
We find that a model with generalized exponential profile and exponent
$n \simeq 1.2$ fits the surface density distribution equally well or even 
slightly better. 
Thus it is possible that the density profile of the galaxy continues 
further outwards and that its true limit lies at densities that are below 
the current observational noise level. \\
 
In the light of these results, Piatek et al.'s recent evidence for Draco 
stars immediately beyond the border of the standard ellipse of semi-major 
axis 28\farcm3 (fields E1, W1, SE1 of Piatek et al.\ 2001) has to be 
understood as a detection of the ordinary Draco population and is not an 
indication of an extratidal component.
Piatek et al.\ claim that there is also weak evidence for Draco stars in 
their fields E2 and W2, i.e.\ at angular distances of about $1.3^\circ$
from the center.  
However they say that one may disagree with this view because their data 
are such that they can neither prove nor disprove the presence of Draco 
stars at these larger radii. Our results indeed suggest that there is 
no Draco component at large radii down to surface densities of 
$10^{-3}$ of the central density.    
In the elliptical annulus with semi-major axis radii between $40'$ and $60'$ 
we find mean surface densities above background that are 
$\le 6\times 10^{-4}$ (S1) and $\le 1\times 10^{-3}$ (S2) of the central 
density and compatible with zero within the statistical error. 
IH95 however reported densities above background of the order of 
$10^{-2}$ times the central density in the distance range from 
$40'$ to $70'$ (cf.\ panel 2 of their Fig.\,2). 
This result has often been regarded as a major piece of evidence for the 
existence of an extratidal Draco component. 
With a limiting magnitude of $R \simeq 21$ the IH95 study of Draco does 
not comprise fainter stars than our SDSS samples. On the other
hand, above the limit of $i^*=21.7$ our sample S1 is unlikely to miss a 
substantial fraction of the Draco population since the filter mask is 
sufficiently wide (see Fig.~3c) to include all relevant parts of the 
color-magnitude diagram. Yet, the color-magnitude filter provides us with 
the advantage of a much higher contrast. Our counts of specifically 
``Draco-like'' stars thus prove that surface densities as high as those 
reported by IH95, which are ten times higher than in our samples, cannot be 
due to Draco members. 
An extratidal Draco component with a surface density as given by IH95 is 
clearly ruled out. 
Instead, we attribute the counts of IH95 beyond $40'$ to residual foreground 
field stars. \\

The results of the study by SKH97, which also favored an extratidal
Draco component, have turned out to be at least doubtful. 
A detailed comparison, for which mean densities were determined in the same 
fields as observed by SKH97, revealed that their detection of overdensities 
may be due to an accidental underestimation of the density of foreground 
stars and thus spurious. We find that the field that was used to set 
the foreground reference in SKH97 has an extraordinarily low surface density 
of ``Draco-like'' stars which is not representative of other and more distant 
fields. With respect to the mean density of stars at more than 
$2^\circ$ distance from Draco we do not find significant overdensities 
in the SKH97 fields.  \\

In total, there remains no convincing evidence for an extratidal Draco 
component. This however does not necessarily mean that such a component 
is completely absent. For the magnitude range down to $i^*=21.7$, i.e.\ 
the giant branch, horizontal branch and subgiant branch, our study 
yields upper density limits of $3\times 10^{-3}$ times the central 
density (from the two-dimensional distributions) and of $10^{-3}$ times 
the central density (from the radial profiles). The latter limit 
corresponds to a surface brightness of about 33.5~mag\,arcsec$^{-2}$.   
In a recent numerical study of the dynamical evolution of dwarf galaxies 
in a Milky Way halo potential Mayer et al.\ (2001) found tidal 
streams with a maximum surface brightness of 30~mag\,arcsec$^{-2}$. 
An extratidal Draco component on this level of surface brightness would 
have shown up in our data and can therefore be ruled out.  
Extratidal extensions with surface brightness below the above given limits
may however exist. 
N-body simulations of the tidal disruption of dwarf galaxies 
by Oh, Lin \& Aarseth (1995) have indeed produced tidal tails with surface 
densities between $10^{-4}$ and $10^{-3}$ of the central density of the dwarf.
Thus it may be that despite the significant improvement in surface density 
contrast our maps of Draco are not yet sensitive enough to reach the level 
at which tidally induced extensions turn up. At magnitudes beyond our 
cut-off, nothing is known about signs of tidal perturbation because no 
observations are available so far except for the very center of Draco. 
Since one does not expect mass segregation to be present in such a 
low-density system like Draco tidal effects should in this case not depend 
on the brightness of the tracer population. Therefore the above given 
limit in relative surface density should hold also for fainter stars. 
However, fainter stars, i.e.\ in the region of the main-sequence turn-off 
and below, have the advantage of being much more numerous than those on 
the giant and subgiant branch (see the luminosity function Fig.~10b). 
On the other hand the number of field stars 
with colors in the range of the Draco turn-off will not increase 
accordingly, because the contaminating stars in this part of the CMD 
will be halo turn-off and main sequence stars at about the same 
distance as Draco. 
Thus, from the point of number statistics and contrast, Draco stars at and 
beyond $i^* = 23$ mag appear to be much better suited for an eventual 
detection of an extratidal component. The goal of future wide-field studies 
of Draco must therefore be to push the magnitude limit down to below the 
main sequence turn-off. \\

Since there is no evidence for an extratidal component around Draco 
(down to the above limit of surface density) the question arises why 
other Milky Way dSph's such as Sagittarius, Ursa Major and Carina  
do show tidal effects (see \S 1) while Draco does not. The clearest 
evidence of tidal perturbation undoubtedly exists for the Sagittarius 
dwarf. The case of Sagittarius is however not comparable to any of the 
other known Milky Way companions since it is much closer to the Galactic 
center ($R\simeq 19$~kpc for Sagittarius, $R\ge 66$~kpc for other MW dSph's). 
Thus for Sagittarius one can directly infer from its current position 
that it must be subject to strong tidal forces whereas this is not certain 
for other more distant dwarf spheroidals as long as their orbits are 
unknown. In other words, it is by no means clear whether 
Draco or any of the other dSph's has penetrated as deeply into the Galactic 
potential during its history as we presently observe it for the Sagittarius.
The Ursa Major dSph is comparable to Draco in that its present galactocentric
distance ($R \simeq 66$~kpc, Mateo 1998) is similar to that of Draco, 
that it has a very low luminosity, and that a very high mass-to-light ratio 
($M/L \simeq 80$, Mateo 1998) is obtained via conventional mass estimation 
based on the central velocity dispersion. 
On the other hand, the morphology of the Ursa Major dSph is quite different 
from that of Draco because it is much more flattened ($e=0.56$, IH95) and 
reveals a clumpy internal structure (IH95; Kleyna et al.\ 1998). 
These peculiarities suggest that the Ursa Major dSph has undergone much 
stronger tidal interactions with the Galaxy than Draco. It is then not so 
surprising that Ursa Major (in contrast to Draco) has a supposedly extratidal
halo with a surface density of about $10^{-2}$ of the central surface density
as found by Martinez-Delgado et al.\ (2001). 
The reason might be that Ursa Major is on a galactic orbit that has smaller 
pericentric distance than the orbit of Draco. In this case it would however 
be unlikely that Draco and Ursa Major belong to the same great circle stream 
as proposed by Lynden-Bell \& Lynden-Bell (1995). 
The key to answering these questions lies in precise measurements of the 
tangential velocities (see end of \S~6.2). 
The case of the Carina dwarf spheroidal is difficult to understand as long
as no information on its orbit is available. With a galactocentric 
distance of about 100~kpc (Mateo 1998) it is presently much more distant 
than Draco, so strong tidal perturbations seem a priori less likely.  
On the other hand, Majewski et al.\ (2000) have identified an extended 
distribution of Carina-like giants with a surface density on the level of 
about $10^{-2}$ times the central surface density and a shallow radial 
profile which is thought to be rather typical of an extratidal component. 
The fact that Carina's total mass as estimated in the usual way from 
its velocity dispersion and its surface density profile 
($M/M_\odot \simeq 1.3\times 10^7$, Mateo 1998)
is a factor of two to three lower than the likewise estimated total mass 
of Draco is not sufficient to explain the apparently very different amounts 
of tidal mass loss from these objects. One may wonder whether the remarkably 
different star formation histories of Carina and Draco perhaps indicates 
fundamental differences in the evolution of the two galaxies that could
have an impact also on the existence of an extratidal component.

\subsection{Draco's structure and basic parameters}

Apart from setting a new and stringent upper limit on the surface density 
of extratidal stars around Draco, the SDSS data provide us with improved 
knowledge of the structure and the basic parameters of Draco's main body. 
We find that its spatial extent and projected shape is essentially that of
an ellipsoid with a core radius of about 160~pc (referring to a decline 
to about 26\% of the central surface density), an outer radius of at least 
820~pc (both along major axis), and a radius-independent ellipticity of 
0.29. There is possibly a 10\% deviation from this simple symmetric structure 
in the outer part. 
The simplicity of the spatial structure of the Draco dSph is a very 
important fact, as is the lack of evidence for an extratidal component 
around it, because both findings lend support to the fundamental assumption 
that the system is in dynamical equilibrium or at least close to it. 
This is one of the key assumptions generally made in order to estimate the 
dynamical masses of systems like the dwarf spheroidals. 
The shape of Draco's radial density profile further suggests that a King 
model should indeed be an acceptable approximation of the system.     
Under these assumptions and in combination with kinematic data the spatial 
characteristics of the system imply that it has a total mass of 
2.2 - 3.5$\times 10^7~M_\odot$. If the distribution of mass in Draco is 
less spatially concentrated than is indicated by its luminous matter, the 
total mass out to $r_t$ could be by up to a factor of 45 higher.  
To check whether the equilibrium hypothesis and the dynamical model is 
really adequate one would need detailed information on the velocity profile 
of Draco, which is currently not known. A more definitive answer on the 
question of the total mass of Draco must therefore still await future 
detailed kinematic measurements.\\

The luminosity of Draco has been determined in a straightforward and 
model-independent way by integrating the observed flux on Draco's
giant branch, horizontal branch, and subgiant branch over the complete 
volume of the galaxy. 
For the residual luminosity from Draco stars beyond the magnitude limit of 
the SDSS sample we also obtained a direct and almost assumption-free estimate 
using supplementary Draco data from HST. The resulting total magnitude
of $m_{i^*} = 10.50$ covers the complete range from the brightest 
giants at $i^*=16.0$ down to the presumed maximum of the luminosity 
function at $i^* = 28.1$. Thus there remains little room for 
significant deficits due to still unaccounted contributions of light. 
The total $i$ band luminosity is $(L/L_\odot)_{i} = 2.4\times 10^5$, 
clearly demonstrating that the Draco dSph is substantially less luminous 
than the brightest globular clusters in our Galaxy although it is at least 
a factor of 10 more spatially extended (in the sense of its linear diameter) 
than those clusters. \\

The low luminosity is in obvious contrast with the estimated total mass of
Draco. A total mass of 2.2 - 3.5$\times 10^7 M_\odot$ yields a global 
mass-to-light ratio of 92 - 146~$(M/L_i)_\odot$ with respect to the $i$ band, 
which lies at least 1.5 orders of magnitude above the normal values of the 
mass-to-light in systems with similar content of luminous matter, namely 
globular clusters ( 1 - 3 $(M/L)_\odot$). In the compilation of Mateo 
(1998, Table 4) the only other Local Group dwarf galaxy with a similarly 
high $M/L$ is Ursa Minor, while most other dSphs have estimated 
$M/L$ ratios $< 40$. Taken at face value, the above mass-to-light ratio
necessarily implies that Draco must be dominated by some sort of dark 
matter component. Even if our mass estimate were wrong by a factor of
2 - 5, the conclusion that a large fraction of the mass in Draco must come 
from a dark component is inescapable. A deviation from the assumption 
that ``mass follows light'', in the sense that the dark matter component 
in Draco could -- as in other galaxies -- be less concentrated than the 
visible matter, would further increase the mass-to-light ratio. 

Hence there exist essentially two possibilities: (a) 
The conventional mass estimates must deviate strongly from the true mass
so that the real mass-to-light ratio is normal. This could be the case 
if the system is very far from equilibrium or unbound.
(b) The Draco dwarf must contain a large fraction of its mass as dark 
matter, whatever the exact quantity of this component may be.
The fact that the spatial structure of the Draco population does not reveal 
any sign of a perturbation of its internal dynamics makes the first 
possibility unlikely and points towards the second one.\\

Using a simple logarithmic halo potential we showed that constraints 
from the Galactic tidal field strongly support this conclusion.  
With a normal mass-to-light ratio, which means a total mass of 
not more than $7\times 10^5 M_\odot$, the observed size of Draco would 
clearly exceed the Lagrange limit for any reasonable value of the 
galactocentric distance. Hence the outer part of Draco should be
heavily affected by tidal stripping, which is not observed. 
How much higher the mass-to-light ratio has to be depends on how closely 
Draco's orbit approaches the Milky Way. Taking twice the value of the 
radial velocity as a guess for the unknown tangential velocity   
yields a hypothetical pericentric distance of 48~kpc.  
To reach consistency between the measured radius $r_t$ and the Lagrange 
limit down to this pericentric distance a mass-to-light ratio
of $M/L \approx 30$ is required. Values near $M/L = 100$ would enable 
Draco to maintain its current size even at pericentric distances of 
25~kpc. Whether Draco is likely to come that close to the Galaxy 
can only be found out with very precise measurements of Draco's 
absolute proper motion. The tangential velocity would in this case
be about 120~\kms and the (galactocentric) proper motion about 
0.4~mas/a. The future astrometric satellite missions SIM and GAIA 
are expected to measure this motion with high accuracy so that 
firm constraints on Draco's orbit and hence on the theoretical tidal 
boundary can then be obtained. 
  
This paper has focussed on analysing the properties of the 
Draco dwarf spheroidal galaxy as a whole. Apart from this the SDSS data 
also enable us to investigate the stellar populations of Draco 
(i.e., red giants, AGB stars, subgiants, red and blue horizontal branch 
stars, and RR Lyrae variables) individually and to compare their 
distribution. 
This study is currently under way and will be described in 
a future paper (Odenkirchen et al. 2001, in preparation). \\

\acknowledgments

We thank S.\ Piatek and an anonymous second referee for comments and 
suggestions which helped to improve this paper. \\ 
The Sloan Digital Sky Survey (SDSS) is a joint project of The University of 
Chicago, Fermilab, the Institute for Advanced Study, the Japan Participation 
Group, The Johns Hopkins University, the Max-Planck-Institute for Astronomy 
(MPIA), the Max-Planck-Institute for Astrophysics (MPA), New Mexico State 
University, Princeton University, the United States Naval Observatory, and 
the University of Washington. Apache Point Observatory, site of the SDSS 
telescopes, is operated by the Astrophysical Research Consortium (ARC). 
Funding for the project has been provided by the Alfred P. Sloan Foundation, 
the SDSS member institutions, the National Aeronautics and Space 
Administration, the National Science Foundation, the U.S. Department of 
Energy, the Japanese Monbukagakusho, and the Max Planck Society. 
The SDSS Web site is \url{http://www.sdss.org/}. \\
Part of this work is based on observations with the NASA/ESA Hubble 
Space Telescope, obtained from the data archive at the Space Telescope 
Science Institute, which is operated by the Association of Universities 
for Research in Astronomy, Inc. under NASA contract No. NAS5-26555.

%% Remarks for AAS manuscripts

%% Generally speaking, only the figure captions, and not the figures
%% themselves, are included in electronic manuscript submissions.
%% Use \figcaption to format your figure captions. They should begin on a
%% new page.

%% No more than seven \figcaption commands are allowed per page,
%% so if you have more than seven captions, insert a \clearpage
%% after every seventh one.

%% There must be a \figcaption command for each legend. Key the text of the
%% legend and the optional \label in curly braces. If you wish, you may
%% include the name of the corresponding figure file in square brackets.
%% The label is for identification purposes only. It will not insert the
%% figures themselves into the document.
%% If you want to include your art in the paper, use \plotone.
%% Refer to the on-line documentation for details.

\clearpage

%% Tables should be submitted one per page, so put a \clearpage before
%% each one.

%% Two options are available to the author for producing tables:  the
%% deluxetable environment provided by the AASTeX package or the LaTeX
%% table environment.  Use of deluxetable is preferred.
%%

%% Three table samples follow, two marked up in the deluxetable environment,
%% one marked up as a LaTeX table.

%% In this first example, note that the \tabletypesize{}
%% command has been used to reduce the font size of the table.
%% Note also that the \label command needs to be placed 
%% inside the \tablecaption.

\begin{figure*}[t]
\includegraphics[scale=1.0]{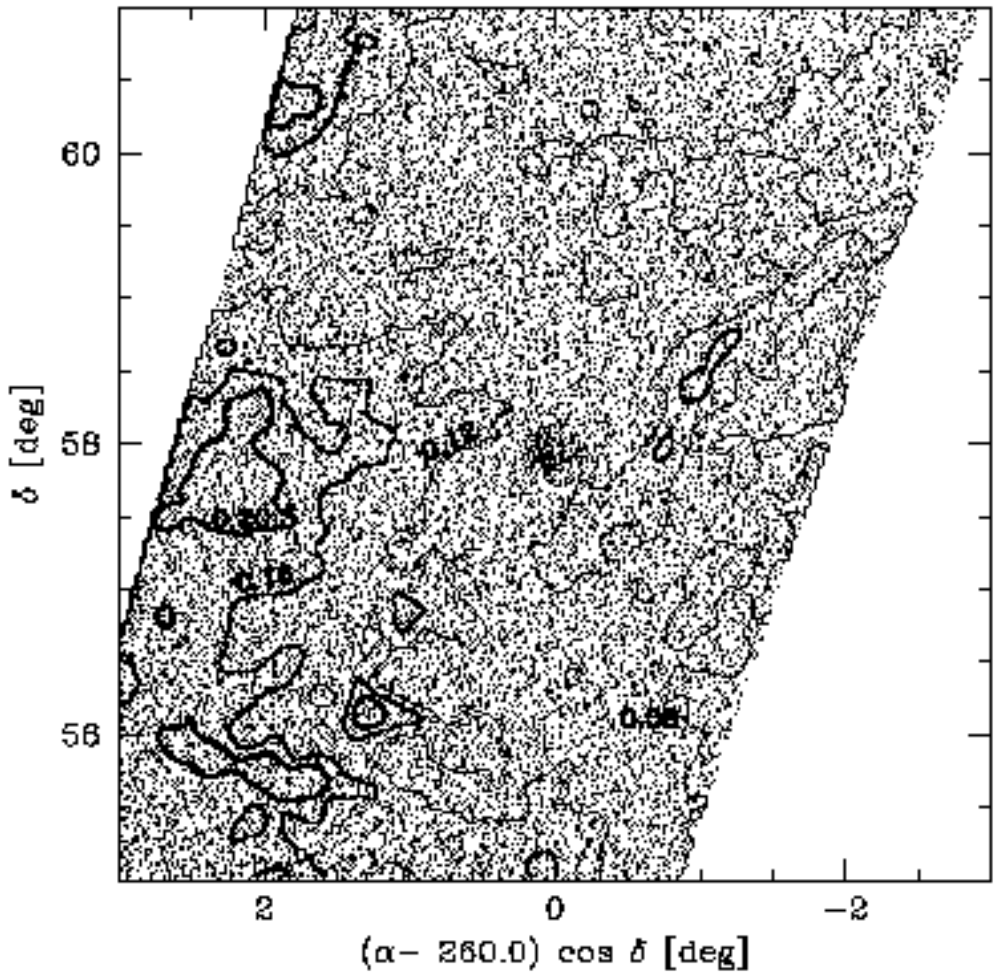}
\figcaption[fig1.eps]{Spatial distribution of SDSS point sources
(dots) and interstellar extinction (contours) in the region of the Draco 
dwarf spheroidal galaxy. The concentration of dots in the center of the 
field reveals the dwarf spheroidal. In order to avoid overcrowding only 
1/10 (!) of the total sample of point sources is plotted. The contours 
indicate extinction levels of 0.08, 0.12, 0.16, and 0.20 mag in $A_{g^*}$
in the order of increasing line thickness. Extinctions in $u$, $r$, 
$i$, and $z$ are obtained by scaling the values of $A_{g}$ with 1.36, 
0.73, 0.55, and 0.39, respectively.   The extinction values are 
based on the reddening maps of Schlegel et al.\ (1998).    
\label{fig1}}
\end{figure*}

\begin{figure*}[t]
\includegraphics[scale=0.95]{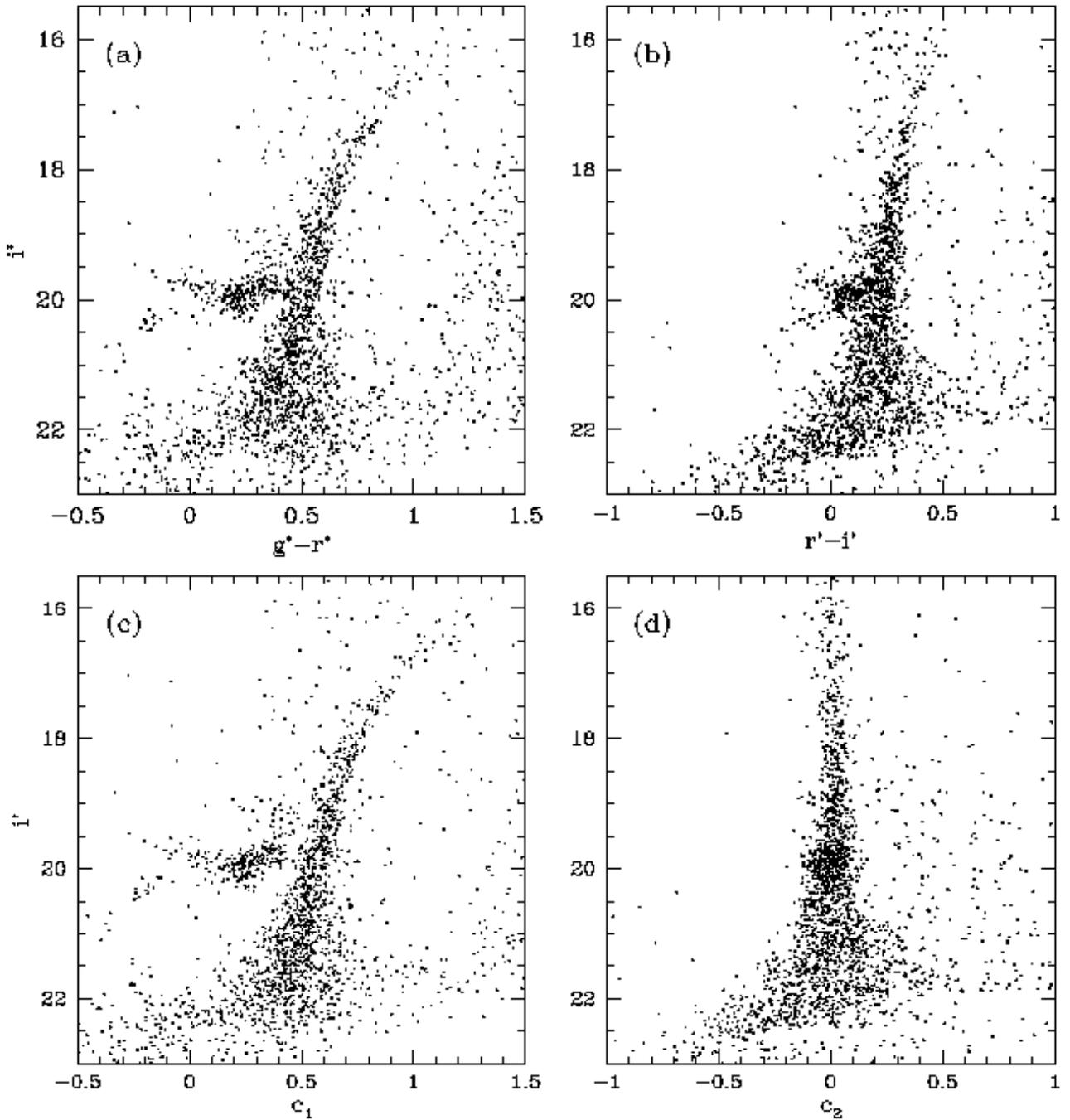}
\figcaption[fig2.eps]{Color magnitude diagrams of SDSS stars in the 
core region of the Draco dSph (ellipse with semi-major axis of $9'$
and ellipticity 0.3 centered on $(\alpha,\delta) = 
(260.056^\circ,+57.915^\circ)$) using different color indices. The diagram
of $i^*$ versus $c_1$ (panel (c)) optimizes the distinction of 
the different types of Draco stars, i.e., giant branch, asymptotic 
giant branch, subgiant branch, red horizontal branch, RR Lyrae 
stars and blue horizontal branch. 
In the ``orthogonal'' color index $c_2$ (panel (d)) Draco stars of all 
magnitudes scatter around zero. 
For the definition of the indices $c_1$ and $c_2$, see Eq.1.    
\label{fig2}}
\end{figure*}

\clearpage

\begin{figure*}[t]
\includegraphics[scale=0.87,bb=28 395 560 580,clip=true]{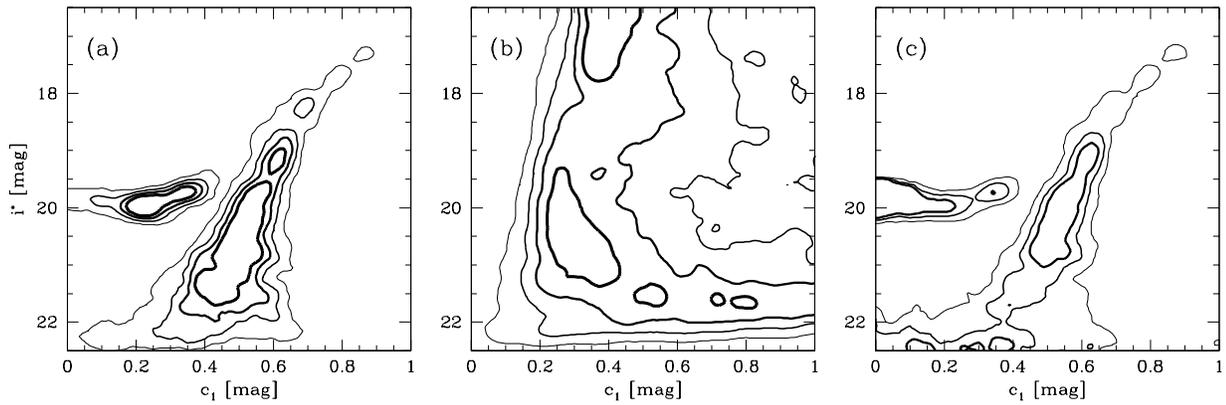}
\figcaption[fig3.eps]{Contour plots showing the distribution 
of stars in the color-magitude plane of $(c_1,i^*)$. (a): Density 
distribution $f_D$ of Draco stars from counts in an ellipse of semi-major 
axis $30'$, ellipticity 0.3, and position angle $90^\circ$, centered on 
the Draco dSph (contribution from field stars subtracted). 
(b): Density distribution $f_F$ of field stars from counts at $\ge 2^\circ$ 
angular distance from the center of Draco. (c): Lines of constant number 
ratio $s = f_D/f_F$. In all plots the contour levels are equidistant and 
increase with line thickness. 
\label{fig3}}
\end{figure*}

\clearpage

\begin{figure*}[t]
\begin{center}
\includegraphics[scale=1.0]{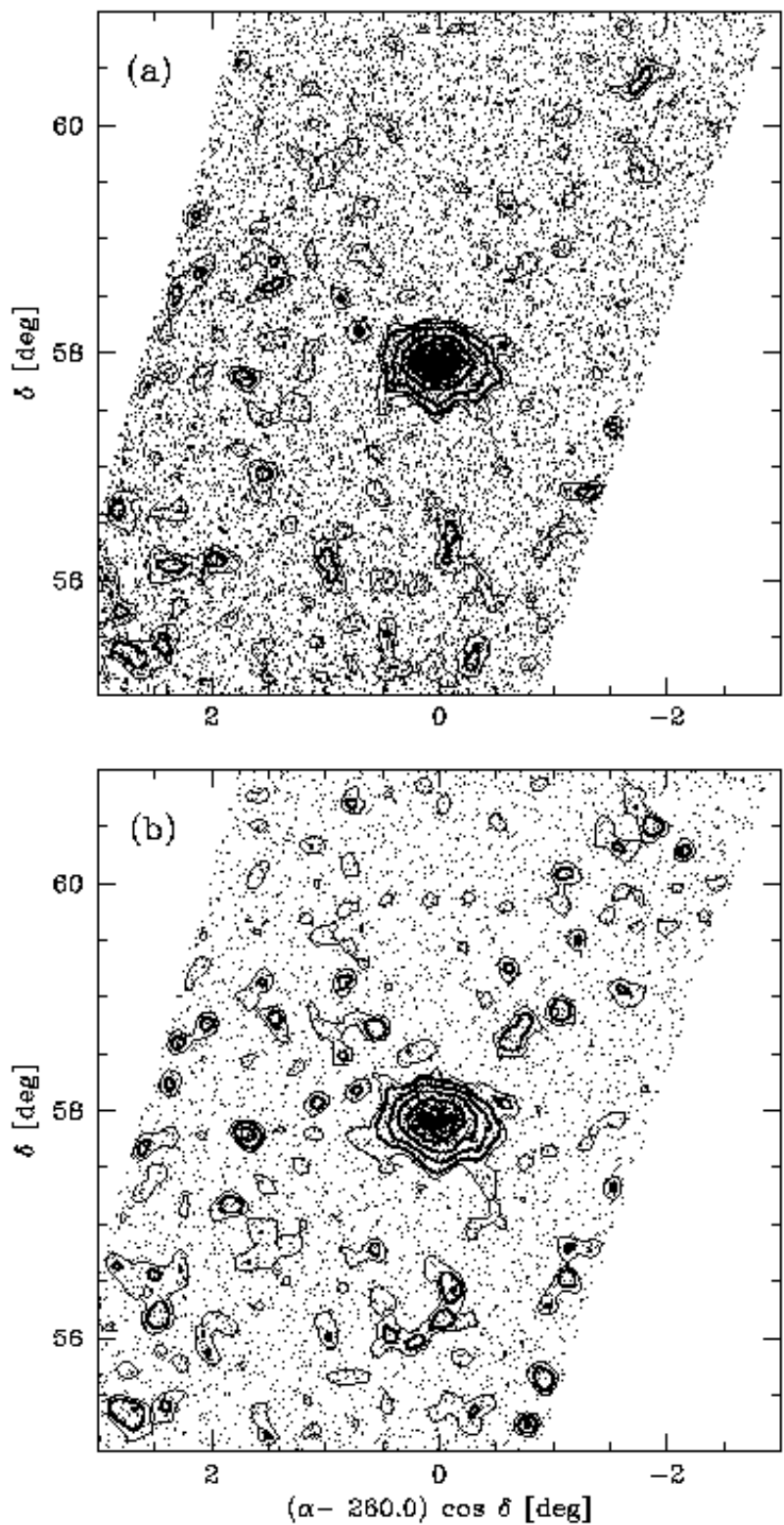}
\figcaption[fig4.eps]{Spatial distribution of photometrically
selected Draco candidates. Dots represent individual stars, lines are 
contours of equal stellar surface density. 
The thin line shows the level of $1\sigma$ above mean background density, 
bold lines show levels of $2\sigma$, $5\sigma$ and higher 
($\sigma$ = rms background variation).
(a): Sample S1, i.e., stars lying within the outermost contour line of 
Fig.~3c. Contour levels are 0.185, 0.22, 0.31, 0.6, and 1.6 arcmin$^{-2}$. 
The mean density of the background is 0.154 arcmin$^{-2}$.
(b): Sample S2, i.e., stars lying within the middle contour of Fig.\,2c.
Contour levels are 0.097, 0.12, 0.18, 0.3, and 0.8 arcmin$^{-2}$. The mean 
density of the background is 0.076 arcmin$^{-2}$. 
\label{fig4}}
\end{center}
\end{figure*}

\clearpage

\begin{figure*}[t]
\begin{center}
\includegraphics[scale=0.9,bb=63 440 550 680,clip=true]{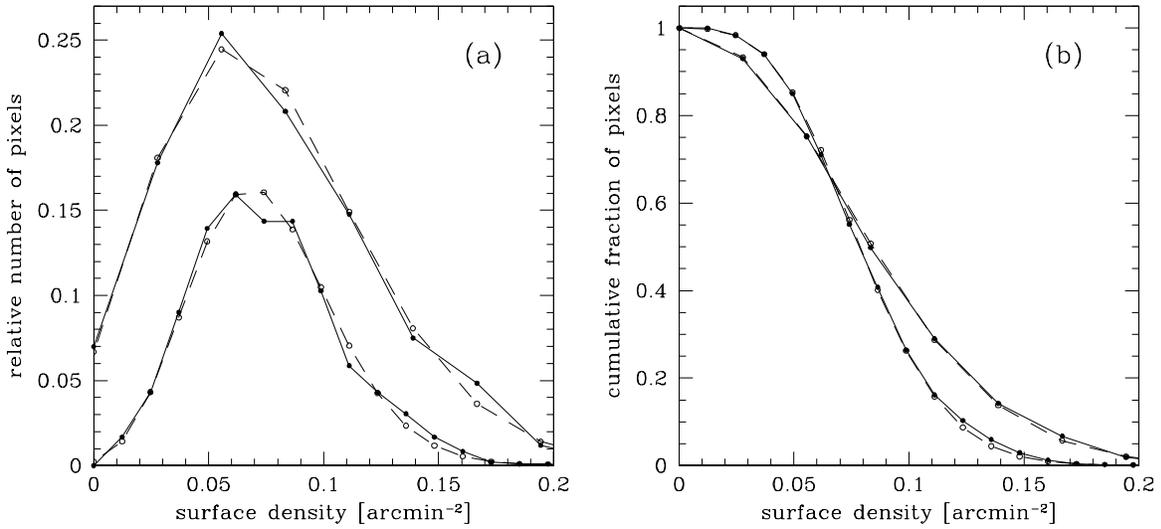}
\figcaption[fig5.eps]{Statistics of the local surface density of 
field stars in the sample S2. (a) Normalized histograms of density values 
measured in pixels of $6'\times 6'$ (upper solid curve) and 
$9'\times9'$ (lower solid curve) at angular distances of more than 
$45'$ from the center of Draco. For comparison, Poisson distributions 
for the observed mean values are drawn (dashed lines).  
(b) Cumulative distributions corresponding to the histograms shown in (a). 
The fraction of pixels with densities above a given threshold is seen
to agree with predictions for Poissonian fluctuations.
\label{fig5}}
\end{center}
\end{figure*}

\clearpage

\begin{figure*}[t]
\begin{center}
\includegraphics[scale=1.0]{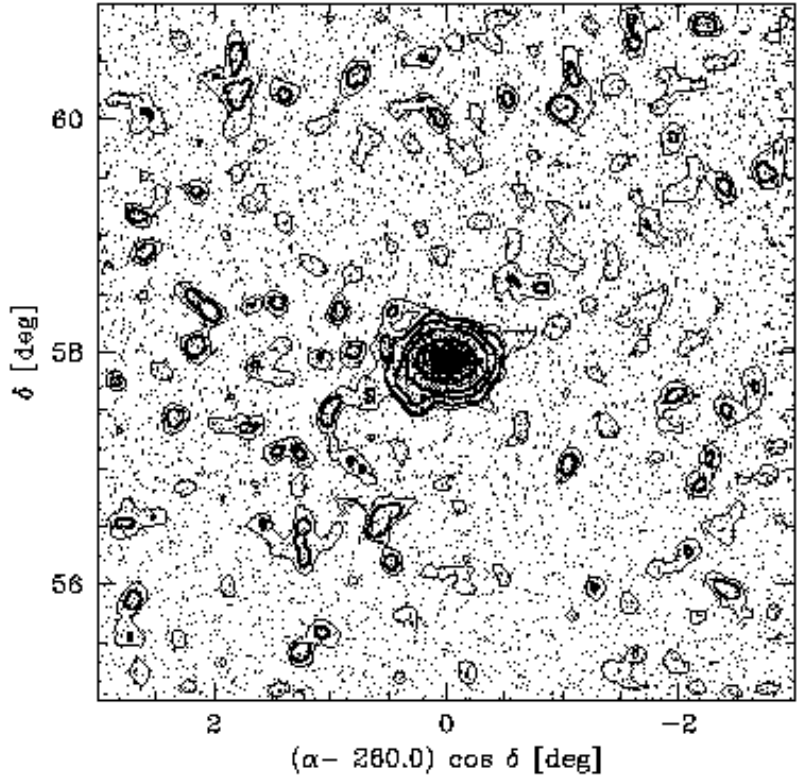}
\figcaption[fig6.eps]{Monte-Carlo simulation of a star field 
with the same basic parameters as sample S2. The background has constant 
mean surface density of 0.076~arcmin$^{-2}$, the Draco galaxy is modelled 
as an ellisoid with a generalized exponential profile 
(see \S 4 and Table 2). The surface density has been sampled 
and smoothed in exactly the same way as in the case of the observed 
samples shown in Fig.~4. The contour level are the same as those in 
panel (b) of Fig.~4.
\label{fig6}}
\end{center}
\end{figure*}

\clearpage

\begin{figure*}[t]
\includegraphics[scale=0.95,bb=50 430 535 675,clip=true]{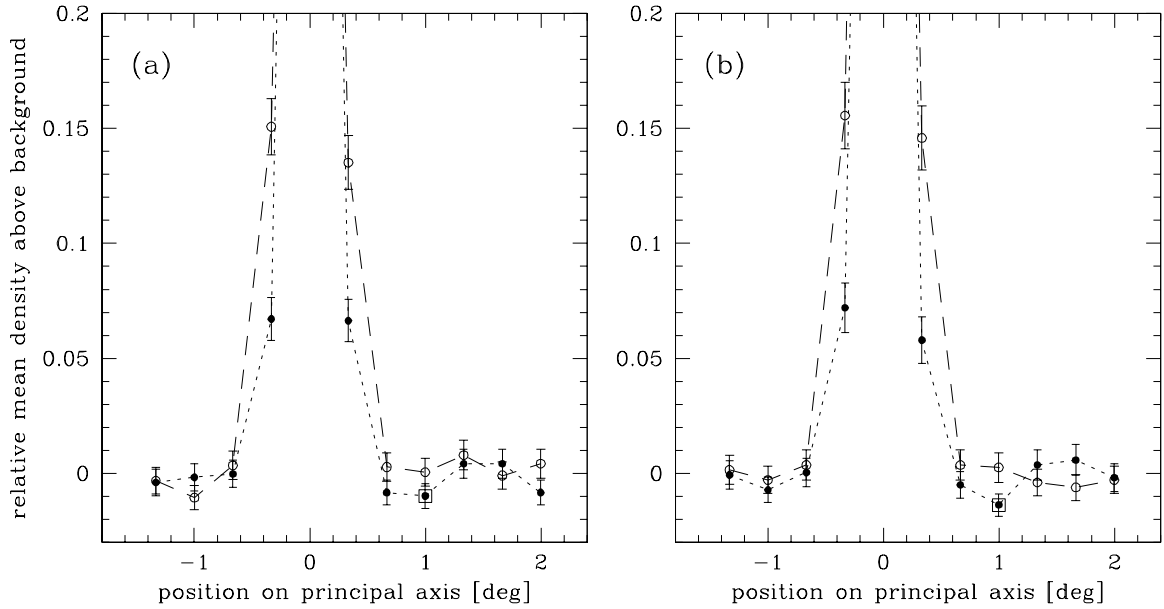}
\figcaption[fig7.eps]{Relative mean surface density in subfields 
of $23' \times 23'$ placed along the conventional principal 
axes of the Draco galaxy.  
The densities are scaled to the mean density above background in the 
central subfield. The mean background density, determined from stars 
at more than $2^\circ$ distance from Draco, is subtracted. 
Panel (a) is for sample S1, panel (b) for sample S2. 
Open circles show the densities for the fields along the major axis 
(position angle $82^\circ$), filled circles are for the fields along 
the minor axis. The angles along these axes increase from south to north 
and from east to west, their zero point is in the center of Draco. 
These plots are designed such that they allow direct comparison with 
Fig.\,2 of Smith et al.\ (1997, SKM97). The field that was used as the
foreground reference in SKM97 ($1^\circ$ north of Draco) is marked by 
an open square (for details see text). 
\label{fig7}}
\end{figure*}

\clearpage

\begin{figure*}[t]
\begin{center}
\includegraphics[scale=0.9,bb=63 200 550 680,clip=true]{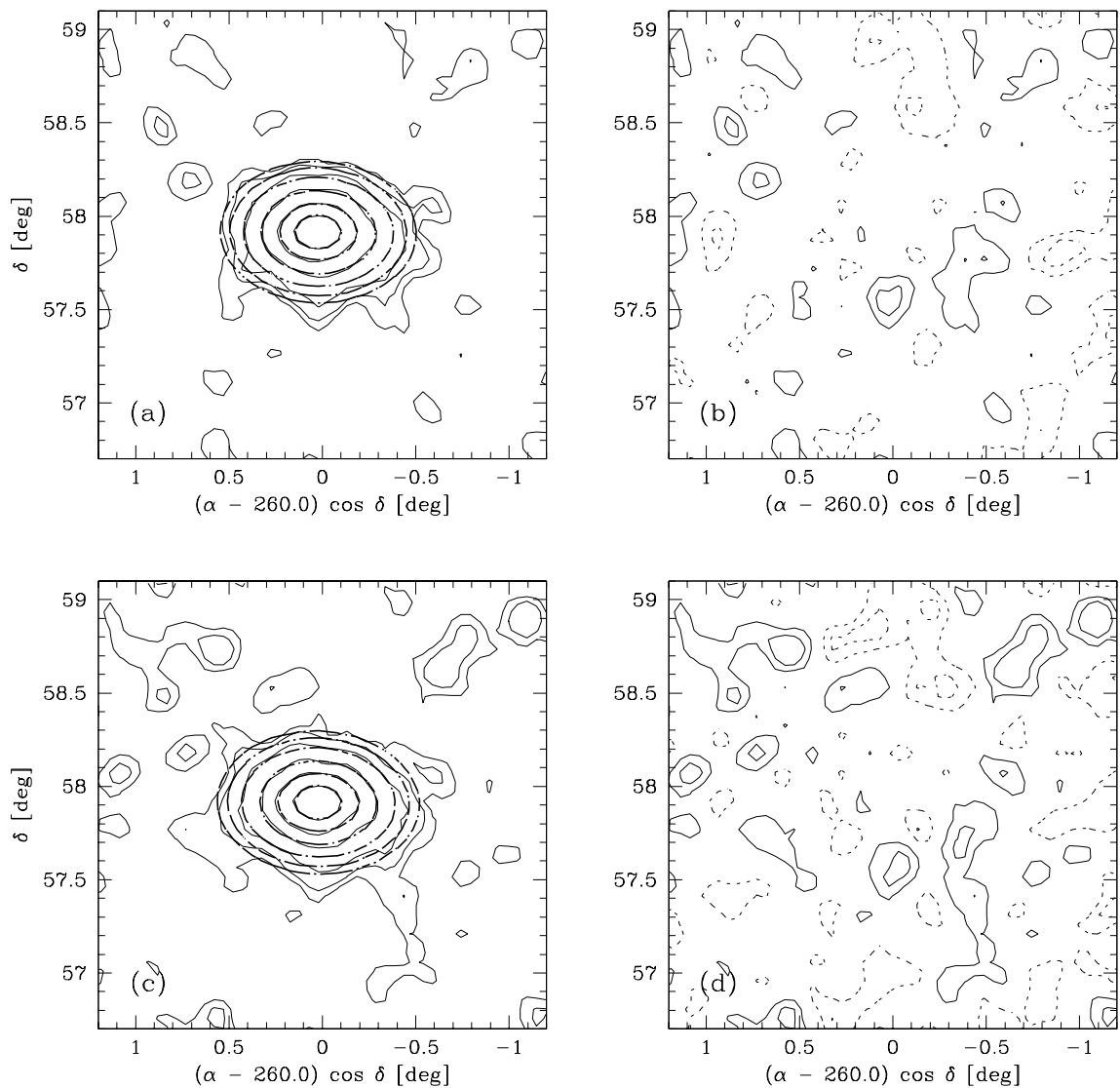}
\figcaption[fig8.eps]{
Contour plots of the observed distribution of stars of Draco, the 
corresponding best-fit elliptical model distribution, and the residuals of 
this fit. (a) Surface density contours for sample S1: 
Thin solid lines show the observed surface density (the same as in Fig.~4). 
The three outermost contours are drawn at levels of $1\,\sigma$, $2\,\sigma$ 
and $5\,\sigma$ above background, $\sigma$ being the rms of the background 
variations. 
Thick dot-dashed lines show the contours of the best-fit exponential model 
(Sersic profile with $n=1.2$) at the same levels.  
(b) Residuals $(O-C)$ of the fit shown in panel (a), rescaled to the level 
of the mean background counts by means of the square root of the local 
amplitude of the model. Solid lines show positive residuals, dashed lines 
show negative ones. The contour levels correspond to $\pm 1\sigma$ and 
$\pm 2\sigma$. (c): Same as (a), but for sample S2. (d): Same as (b), but 
for S2. For details on the samples and the models, see text.    
\label{fig8} }
\end{center}
\end{figure*}

\clearpage

\begin{figure*}[t]
\begin{center}
\includegraphics[scale=0.9,bb=65 220 380 680,clip=true]{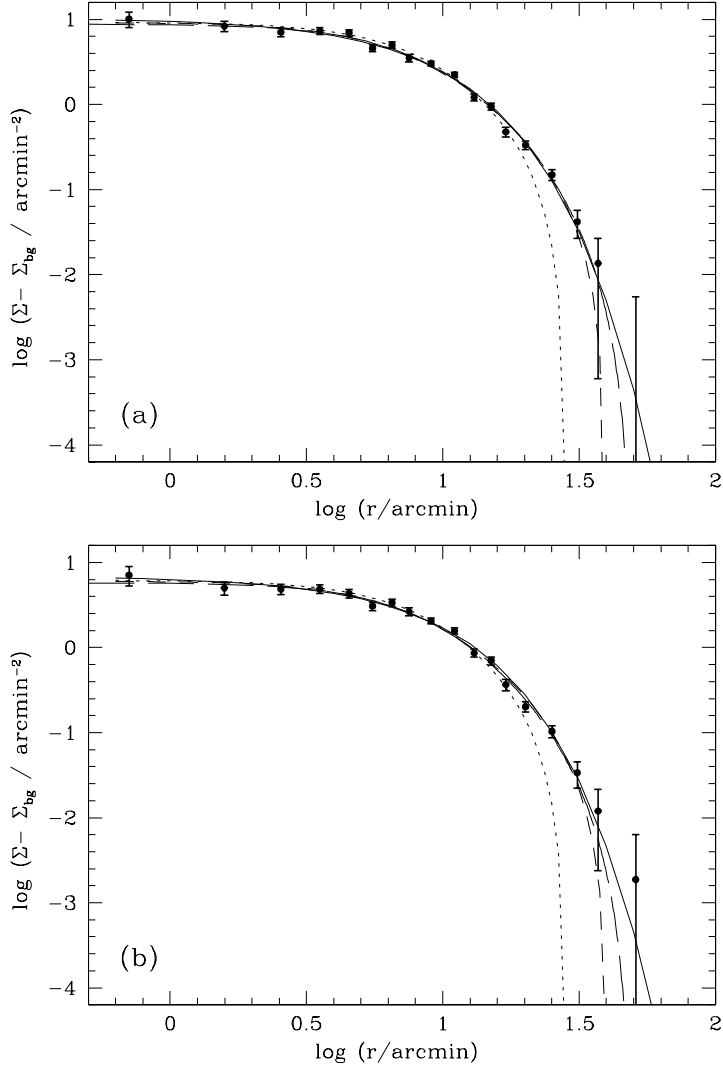}
\figcaption[fig9.eps]{
Radial profiles of the surface density of stars in Draco. 
Data points with errorbars show the profile of the observed distribution 
(i.e., star counts in elliptical rings around the center of Draco, 
mean background subtracted). The lines show the profiles of different models,
i.e.\ a K62 model with the parameter values given by IH95 (dotted), 
the new best-fit King models, K62 (short dashed) and K66 (long-dashed), and 
the best-fit Sersic model (solid, see also Fig.~7). 
Panel (a) is for sample S1, panel (b) for sample S2.
\label{fig9} }
\end{center}
\end{figure*}

\clearpage

\begin{figure*}[t]
\begin{center}
\includegraphics[scale=0.9,bb=52 440 550 680,clip=true]{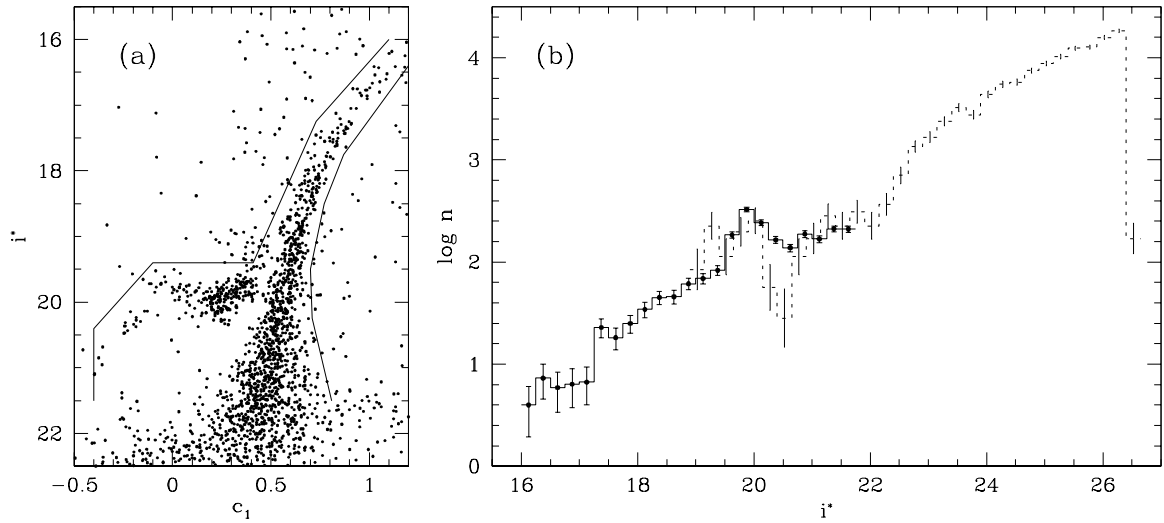}
\figcaption[fig10.eps]{The luminosity function of Draco from SDSS
and HST. 
Panel (a) specifies the region in SDSS color-magnitude space used 
to construct the bright part of the luminosity function. For comparison 
the dots show the stars in the core region of Draco, as in Fig.~1c. 
Panel (b) shows the luminosity function (solid line) that results from 
counting the SDSS stars in the region outlined in (a) within the tidal 
radius of Draco. 
The contribution from field stars has been subtracted. Error bars 
indicate combined $\sqrt{n}$ uncertainties. The dashed line in panel 
(b) shows the luminosity function of fainter Draco stars derived from 
$I$-band HST observations (see text). The magnitude bins are shifted by 
0.4 mag to match the mean offset between $i^*$ and $I$, the counts are 
rescaled such that the integrated numbers in the overlap interval  
($18.9 \le i^* \le 21.7$) coincide. 
\label{fig10}}
\end{center}
\end{figure*}

\begin{figure*}[t]
\begin{center}
\includegraphics[scale=1.0,bb=120 250 410 540,clip=true]{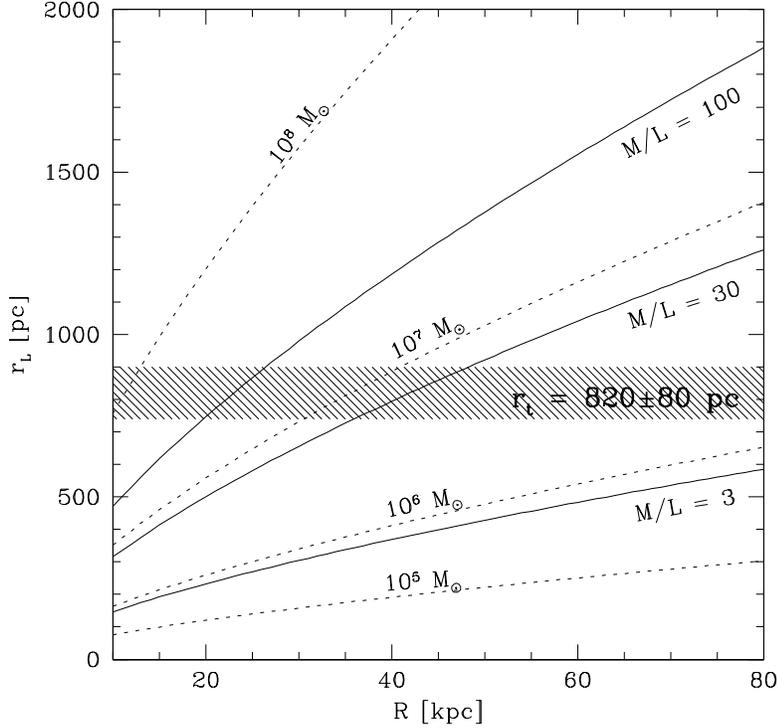}
\figcaption[fig11.eps]{Lagrange radius $r_L$ of the Draco dSph in a 
logarithmic spherical Galactic potential with circular velocity 220~\kms, 
plotted as a function of galactocentric distance $R$ and for 
different satellite masses. 
The solid lines assume the observed luminosity of 
$L= 2.4\times 10^5 L_\odot$ and mass-to-light ratios of 3, 30, and 100.
The shaded region indicates the tidal radius of Draco as determined
by the fit of the King (1962) profile (i.e.\ $r_t=40'$, corresponding to 
820~pc). 
The current galactocentric distance of Draco is $R=71$~kpc.
Its pericentric distance $R_p$ must clearly be smaller, since Draco has 
a galactocentric radial velocity of $v_r = -100$~\kms. 
Assuming, e.g.,  $v_T = 200$~\kms for the unknown tangential velocity 
component, the pericenter would be at $R_p = 48$~kpc.        
\label{fig11}}
\end{center}
\end{figure*}

\end{document}